\documentclass[twocolumn,tighten]{aastex62}

\usepackage{graphicx}
\usepackage{ulem}
\usepackage{amsmath}
\usepackage{wrapfig}

\usepackage{tikz}

\def\Msun{{M_{\odot}}}

\newcommand\lsim{\mathrel{\rlap{\lower4pt\hbox{\hskip1pt$\sim$}}
        \raise1pt\hbox{$<$}}}
\newcommand\gsim{\mathrel{\rlap{\lower4pt\hbox{\hskip1pt$\sim$}}
        \raise1pt\hbox{$>$}}}

\begin{document}
\shorttitle{Planets in Globular Clusters}
\shortauthors{Kremer et al.}

\title{Probing the Survival of Planetary Systems in Globular Clusters with Tidal Disruption Events}

\author[0000-0002-4086-3180]{Kyle Kremer}
\affil{ Department of Physics \& Astronomy and Center for Interdisciplinary Exploration \& Research in Astrophysics (CIERA), Northwestern University, Evanston, IL 60208, USA}

\author[0000-0002-1271-6247]{Daniel J. D'Orazio}
\affil{Department of Astronomy, Harvard University, 60 Garden Street Cambridge, MA 01238, USA}

\author[0000-0003-0607-8741]{Johan Samsing}
\affil{Department of Astrophysical Sciences, Princeton University, Peyton Hall, 4 Ivy Lane, Princeton, NJ 08544, USA}

\author[0000-0002-3680-2684]{Sourav Chatterjee}
\affil{Tata Institute of Fundamental Research, Homi Bhabha Road, Mumbai 400005, India}

\author[0000-0002-7132-418X]{Frederic A. Rasio}
\affil{ Department of Physics \& Astronomy and Center for Interdisciplinary Exploration \& Research in Astrophysics (CIERA), Northwestern University, Evanston, IL 60208, USA}

\begin{abstract}

Among the growing list of confirmed exoplanets, the number of planets identified in dense star clusters remains sparse. Previous analyses have  suggested this may be due in part to dynamical interactions that unbind large fractions of planets from their host stars, limiting the survival of planetary systems in clusters. Thus, alternative detection strategies may be necessary to study planets in clusters that may no longer be bound to a host star. Here, we use the cluster Monte Carlo code \texttt{CMC} to explore the evolution of planetary systems in dense star clusters. Depending on a number of initial conditions, we show that $10-50\%$ of primordial planetary systems are broken through dynamical encounters over a cluster's full lifetime, populating clusters with ``free-floating'' planets. Furthermore, a large number ($30-80\%$) of planets are ejected from their host cluster through strong dynamical encounters and/or tidal loss. Additionally, we show that planets naturally mix with stellar-mass black holes (BHs) in the central regions of their host cluster. As a consequence, up to a few hundreds of planets will be tidally disrupted through close passages of BHs. We show these BH--planet tidal disruption events (TDEs) occur in clusters at a rate of up to $10^{-5}\,\rm{yr}^{-1}$ in a Milky Way-type galaxy. We predict BH--planet TDEs should be detected by upcoming transient surveys such as LSST at a rate of roughly a few events per year. The observed rate of BH--planet TDEs would place new constraints upon the formation and survival of both planetary systems and BHs in dense star clusters. Additionally, depending on various assumptions including the initial number of planets and their orbital properties, we predict typical globular clusters may contain roughly a few dynamically-formed NS--planet systems and up to roughly 100 WD--planet systems. 

\end{abstract}

\section{Introduction}
\label{sec:intro}

It is now generally accepted that planets are ubiquitous in the universe \citep[e.g.][]{Dressing2013,Fressin2013,Petigura2013,Mulders2015,Winn2015}. Despite the fact that one of the first confirmed exoplanets was identified in the Milky Way globular cluster (GC) Messier 4 in orbit with a millisecond pulsar \citep{Lyne1987, McKenna1988,Sigurdsson1993,Thorsett1999}, the growing catalogue of exoplanet detections has shown that planets found in star clusters are far from the norm. Indeed, of the roughly 4,000 confirmed exoplanets to date, only 30 have been observed in star clusters (see, e.g., \citet{Cai2019} for a summary). Furthermore, the PSR B1620-26 b exoplanet still stands alone as the only exoplanet observed in a GC, despite extensive efforts to identify exoplanets in Milky Way GCs such as 47 Tuc \citep[e.g.][]{Gillilon2000,Weldrake2005,Masuda2017}. 

The noticeable lack of observed exoplanets in GCs suggests that environmental processes unique to dense star clusters may limit the survivability, and perhaps even formation, of planetary systems \citep[e.g.,][]{Adams2006,Malmberg2007,Proszkow2009,Spurzem2009,Parker2012,Hao2013,Cai2019}. For example, stellar encounters occurring while a protoplanetary disk is still present may lead to a truncation of the disk, thereby altering or limiting the planet formation process \citep[e.g.,][]{Clarke1993,Ostriker1994,PortegiesZwart2016}. Furthermore, planetary systems that survive this formation stage in a star cluster will be influenced by subsequent weak and strong encounters that will affect the internal dynamics of the planetary systems \citep[e.g.,][]{HurleyShara2002}. Additionally, planet formation is expected to depend in a significant way upon metallicity \citep[e.g.,][]{Fischer2005,Brewer2018}. In low-metallicity environments like GCs, planet formation (particularly giant-planet formation) may be less pronounced compared to high-metallicity environments. Thus, over the past several decades, the prospects for identifying large numbers of exoplanets systems that have survived to the present day in old star clusters have become increasingly bleak on the basis of both observational and theoretical arguments.

However, despite the above-mentioned challenges and the observational null results, it is not clear whether GCs truly are devoid of planets. Most ground-based transit searches for giant planets in star clusters are statistically insignificant \citep[e.g.,][]{vanSadersGaudi2011}. Furthermore, perhaps the observational null results of transit surveys arise simply because only a small fraction of cluster planets are in a transiting architecture. If a large number of planets are unbound from their host star, for example as a consequence of the various dynamical processes relevant in star clusters,  alternative detection strategies are needed.

From a computational perspective, the topic of the dynamical evolution of planetary systems in star clusters has been explored by a number of analyses, particularly in the less massive open-cluster regime. For example, \citet{HurleyShara2002} used direct $N$-body models with up to $2.2\times10^4$ stars to show that planets can be liberated from their host star through dynamical encounters, primarily within the cluster core, creating a population of free-floating single planets in the cluster. Later work by \citet{Spurzem2009} studied the topic with both direct $N$-body models and a hybrid Monte Carlo method to model clusters of up to $3\times10^5$ stars of a single mass and showed similar results, namely that a significant number of planetary orbits will be perturbed (and in some cases, unbound) through close dynamical encounters. More recently, \citet{Chatterjee2012} showed that a large fraction of planetary orbits in massive open clusters similar to NGC 6791 in Kepler's field of view,  a cluster with roughly three orders of magnitude lower central density compared to typical GCs, can remain intact. Even in this case, it may be difficult for Kepler to detect a large fraction of planets due to crowding and their adopted observation strategy \citep{Chatterjee2012}.
See also computational work on the topic of planets in star clusters by, e.g., \citet{Parker2012,Hao2013,Hamers2017,Cai2019}.

Over the past decade, prospects for identifying stellar-mass black hole (BH) populations in GCs have seen a boom in interest. A plethora of observational \citep[e.g.,][]{Maccarone2007,Strader2012,Giesers2018} and theoretical \citep[e.g.,][]{Mackey2007,Morscher2015,Kremer2018c,Askar2018, Weatherford2018,Askar2019, Kremer2019a} evidence now suggests that some clusters retain hundreds to thousands of BHs at present. These BH populations have important implications for the formation of gravitational wave sources \citep[e.g.,][]{Askar2017, Banerjee2017,Hong2018,Fragione2018b,Samsing2018a,Rodriguez2018b}, low-mass X-ray binaries \citep[e.g.,][]{Heinke2005,Ivanova2013,Giesler2018,Kremer2018a}, and also tidal disruption events \citep[TDEs;][]{Perets2016,Lopez2018,Samsing2019,Kremer2019c}.

In this paper, we use our cluster Monte Carlo code \texttt{CMC} to explore the survival of planetary systems in massive and dense GCs with up to $8\times10^5$ stars, core radii of $\lesssim 1\,$pc, realistic cluster initial stellar mass functions, and stellar evolution prescriptions that incorporate our most up-to-date understanding of compact object formation. We show that a large fraction of primordial planetary systems are broken apart through strong dynamical encounters by the time the cluster has a reached a typical present-day age (roughly 12 Gyr). Furthermore, we show that through their interaction with stellar-mass BHs, planets frequently undergo tidal disruption events (TDEs). The concept of stellar-mass-BH--planet TDEs was first discussed in \citet{Perets2016}, which noted that due to the low mass ratio between the two objects, these events should behave qualitatively similar to the widely studied stellar TDEs by supermassive BHs. Here, we present the first study to explore BH--planet TDEs in the context of full-scale cluster models.

The paper is organized as follows: In Section \ref{sec:analytic}, we present simple analytic predictions for the survivability of planetary systems in GCs and the rate of BH--planet TDEs. In Section \ref{sec:models}, we describe the cluster models used in this study and discuss the results from these models in the context of the predictions made in Section \ref{sec:analytic}. In Section \ref{sec:EM}, we discuss the potential detectability of BH--planet TDEs and in Section \ref{sec:other}, we describe several other observational prospects. We discuss our results and conclude in Section \ref{sec:Discussion}.

\section{Analytic motivation}

In this section, we present several analytic estimates of rates of encounters relevant to planets in typical GCs. The goal of this section is to simply motivate the various types of interactions which will be explored in more detail using our full-scale cluster simulations in Section \ref{sec:models}. We begin in Section \ref{sec:dyn_disruption} by discussing the rate at which a planet--star binary is expected to undergo strong binary--single encounters with single stars (non-BHs) within both the core and halo of a GC. We briefly discuss the nature of these encounters (for example, are they impulsive or resonant?) and discuss the implications these encounters have on the overall survival of primordial planet--star binaries. In Section \ref{sec:planetbh}, we go on to discuss the specific case where planet--star binaries interact with stellar-mass BHs and also predict the rate of planet--BH encounters that may lead to a tidal disruption of the planet.
\label{sec:analytic}

\subsection{Dynamical disruption of planetary systems}
\label{sec:dyn_disruption}

A single planet--star binary will undergo strong encounters (where ``strong'' indicates an encounter where the pericenter distance, $r_p$, between the planet--star binary and an incoming single star is of order the planet--star semi-major axis, $a_{\rm{p}}$) with other single stars\footnote{Planet--star binaries will also undergo interactions with other binaries (binary--binary interactions). We focus in this section on the binary--single case for simplicity, but include binary--binary interactions in our cluster simulations described in Section \ref{sec:models}.} in the cluster at a rate given by
\begin{equation}
\label{eq:strong}
\Gamma_{\rm{strong}} = n_{\rm{s}} (\pi a_{\rm{p}}^2) v_\infty \Bigg[1+\frac{2G M_{\rm{tot}}}{a_{\rm{p}} v_\infty^2} \Bigg]
\end{equation}
where $n_{\rm{s}}$ is the number density of single stars (i.e. any object that is \textit{not} a planet or BH in the cluster), $v_\infty$ is the relative velocity of the binary--single pair at infinity, and $M_{\rm{tot}}$ is the total mass of the three-body system.

Adopting values representative of typical GCs ($n_{\rm{s}} = 10^4\,\rm{pc}^{-3}$ and $v_\infty = 10\,\rm{km\,s}^{-1}$) and assuming $a_{\rm{p}}=5\,$AU and $M_{\rm{tot}}=2M_{\odot}$, we predict planet--star binaries within the core of their host cluster undergo strong encounters with other single stars at a rate of roughly $\Gamma_{\rm{strong}}^{\rm{PS-S}} (\rm{core}) \approx 2\,\rm{Gyr}^{-1}$.

In the outer parts of a typical cluster (i.e., at a radial position of twice the half-mass radius), where $n_{\rm{s}} \approx 500\,\rm{pc}^{-3}$ and $v_{\infty} \approx 2\,\rm{km\,s}^{-1}$ are more typical, the rate that a planet--star binary will undergo strong binary--single encounters is roughly $0.35\,\rm{Gyr}^{-1}$. Thus, planetary systems within both a cluster's core and halo will likely undergo multiple strong encounters throughout the lifetime of the cluster. 

Broadly, strong binary--single encounters can be grouped into several categories determined by the value of the relative velocity of $v_\infty$ compared to $v_c$ (the value of $v_\infty$ for which the total energy of the binary--single system is zero; see Equation 1 of \citet{Fregeau2006}) and $v_{\rm{orb}}$ (the orbital velocity of the incoming binary). In the $v_\infty < v_c$ regime, the total energy of the system is negative. These encounters are expected to be resonant, meaning that the encounter will survive for several orbital times \citet{HeggieHut2003}. However, for planetary systems, $v_c$ is generally much lower than the typical cluster velocities ($v_c \lesssim 1\,\rm{km\,s}^{-1}$ compared to $v_\infty \approx 10\,\rm{km\,s}^{-1}$), thus resonant encounters for planetary systems are rare. Instead, planetary systems will more often undergo strong encounters in the impulsive regime, where the total energy is positive and the encounter is better characterized by a single close fly-by. In the impulsive regime, preservation, exchange, and ionization are all possible, the latter two of which will break apart the original planetary system. As described in \citet{Fregeau2006}, for binary--single encounters with non-equal component masses, an intermediate regime where $v_c<v_\infty<v_{\rm{orb}}$ is also possible. Indeed, for GCs with $v_\infty \approx 10 \, \rm{km\,s}^{-1}$ and planet--star binaries with $v_{\rm{orb}} \approx 13 \, \rm{km\,s}^{-1}$(assuming $M_{\rm{s}}=M_{\odot}$, $M_{\rm{p}}=M_{\rm{J}}$, and $a=5\,$AU), this regime may even be most likely.  In this intermediate regime, exchange and ionization are roughly comparable.
We direct the reader to \citet{Fregeau2006} for a detailed discussion of the different outcomes of binary--single interactions with planet components.

As a result of ionizations and exchanges occurring through strong encounters in both a cluster's core and halo, we expect the total number of primordial planet--star binaries to decrease throughout the lifetime of the cluster, with the total number of surviving systems depending upon the orbital parameters (here, the initial semi-major axis) of the planetary systems at birth \cite[see also, e.g.,][]{HurleyShara2002}. Once unbound from their host stars, these ejected single planets will either ``wander'' about the host cluster until undergoing a subsequent dynamical interactions with other objects or, if ejected from their host star with sufficient velocity, will escape from the host cluster entirely \citep{Cai2019}. Even for those single planets retained in the cluster initially after being separated from the host star, the ultimate fate may still be escape from the cluster through mass-segregation \citep[e.g.,][]{HurleyShara2002}. Thus, we expect the total number of planets to decrease throughout the lifetime of a cluster.

We also note that in addition to strong encounters with $r_p$ of order the planet--star semi-major axis, more distant weak encounters may also play a role in the dynamical evolution of planetary systems in GCs \citep[e.g.,][]{Spurzem2009,Cai2017}. In contrast to the strong regime, where the orbital parameters of a binary can be significantly perturbed by a single encounter, the more common weak encounters act through a cumulative process to perturb binary orbits. For example, \citet{Cai2017} noted that in the less-massive open cluster regime ($N \sim 10^4$) where strong encounters are significantly more rare than in GCs, weak encounters may in fact play a dominant role in determining the long-term outcome of planetary systems in a cluster. Furthermore, \citet{Hamers2017} explored the outcomes of strong and weak interactions including tidal effects for planetary systems within a typical GC and showed that the cumulative effect of many weak encounters with tides may be
high-eccentricity migration and the formation of close planet--star binaries (``hot/warm Jupiters''). As noted in \citet{Hamers2017}, at most $1-2\%$ of planetary systems in a typical cluster are converted to hot Jupiters through such mechanisms, therefore this process is unlikely to be important to the total population. In this analysis, we focus our attention on strong encounters, which in typical GCs of masses few$\times10^5\,M_{\odot}$, are expected to be most important from an energy-generating perspective \citep[e.g.,][]{Fregeau2007} and we simply note that, as explored in a number of previous analyses, inclusion of weak encounters may have an additional (albeit small) effect on the survival of planetary systems in GCs. 

\subsection{Interactions of planets with black holes}
\label{sec:planetbh}

One special subset of the strong encounters described by Equation \ref{eq:strong} is the case where the third star incident upon the planet--star system is a stellar-mass BH. As discussed in \citet{Kremer2018a} in the context of forming BH--star binaries, the rate of interaction of BHs with less-massive stellar populations depends not on the total number of BHs in the cluster but on the number of BHs within the region of the clusters where these populations mix and the density of this ``mixing zone'' is related to the total number of BHs in a complex way. 

Assuming a typical stellar BH mass of $20\,M_{\odot}$ and adopting values characteristic of the BH mixing zone (as motivated by simulations performed in previous work such as \citet{Morscher2015} and \citet{Kremer2018a} and as will be confirmed in Figure \ref{fig:radial_dist}), we estimate that planet--star binaries undergo strong binary--single encounters with single BHs at a rate of

\begin{multline}
\label{eq:BHstrong}
\Gamma_{\rm{strong}}^{\rm{PS-B}} \approx 1.5\,\rm{Gyr}^{-1} \Big( \frac{\it{n}_{\rm{BH}}}{10^3\,\rm{pc}^{-3}} \Big) \Big(\frac{\it{a}_{\rm{p}}}{5\,\rm{AU}} \Big) \\ \times  \Big( \frac{\it{M}_{\rm{BH}}}{20\,M_{\odot}} \Big) \Big( \frac{v_\infty}{10\,\rm{km/s}} \Big)^{-1},
\end{multline}where we have assumed gravitational focusing dominates.

Thus, \textit{within the mixing zone}, a planetary system undergoes strong encounters with BHs at a rate comparable to strong encounters with other stars. However, only a small fraction ($\lesssim 10\%$, as will be shown in Section \ref{sec:models}) of all planetary systems are found within the mixing zone, thus throughout the entire cluster, the full population of planetary systems are more likely to undergo strong encounters with non-BHs. 

As planets are ionized from their host stars or exchanged into new binaries with other objects, other combinations of strong encounters involving BHs and planets become possible beyond the planet--star binary plus BH single case considered in Equation \ref{eq:BHstrong} (for example, BH--BH binary plus planet single, BH--planet plus BH single, etc.)
Equation \ref{eq:BHstrong} is intended to simply demonstrate that binary-mediated interactions with BHs and planets are common within a cluster's core. We explore these various other combinations of encounters in more detail in Section \ref{sec:BHplanets}.

\subsubsection{Tidal disruption events}

\begin{figure}
\begin{center}
\includegraphics[width=0.5\textwidth]{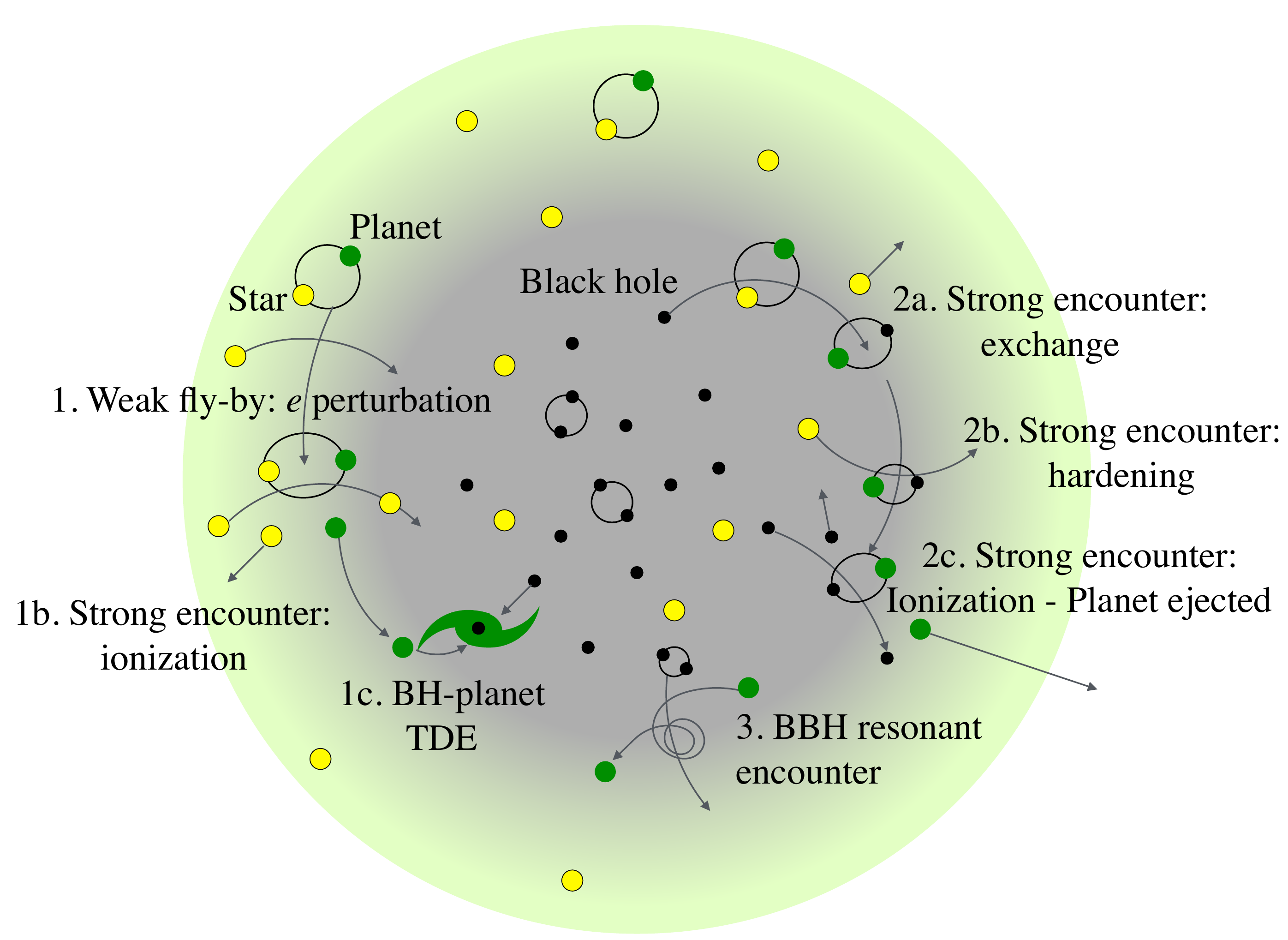}
\caption{\label{fig:cartoon} Cartoon illustration of the dynamical encounters relevant to the evolution of planets in GCs. Shown here are three different outcomes for the dynamical evolution of a typical primordial planet--star system: (1) A planet--star system that is broken apart by an impulsive strong encounter, ultimately leading to a TDE through a single--single encounter with a BH. (2) A planet that exchanges into a binary with a BH companion and is ultimately ejected from the host cluster through a subsequent BH encounter. (3) A single planet that undergoes a resonant encounter with a compact BBH (to be discussed further in Section \ref{sec:BBH-TDEs}).}
\end{center}
\end{figure}

\begin{deluxetable*}{lc|cc|ccc|ccc|cccc}
\tabletypesize{\scriptsize}
\tablewidth{0pt}
\tablecaption{Cluster properties for all model GCs\label{table:models}}
\tablehead{
	\colhead{Model} &
    \colhead{$N_{\rm{P,\,i}}$}&
    \multicolumn{2}{c}{$N_{\rm{P}}\,(\times10^4)$ at 12 Gyr}&
    \multicolumn{3}{c}{$N_{\rm{P}}\,(\times10^4)$ ejected}&
    \multicolumn{3}{c}{BH-P binaries (8-12 Gyr)}&
    \multicolumn{4}{c}{BH--P TDEs}\\
    \colhead{} &
    \colhead{} &
    \colhead{single} &
    \colhead{binary} &
    \colhead{single} &
    \colhead{binary} &
    \colhead{total} &
    \colhead{$N$} &
    \colhead{$\langle n_{\rm{enc}} \rangle$} &
    \colhead{$\langle t_{\rm{lifetime}}/\rm{Myr} \rangle$} &
    \colhead{sin--sin} &
    \colhead{bin--sin} &
    \colhead{bin--bin} &
    \colhead{total}
}
\startdata
1 & $10^5$        &  $3.0$   & $2.1$ & $1.4$ & $3.4$ & $4.8$ & 60 & 3.5 & 24.8 & 35 & 25 & 2 & 62\\
2 & $2\times10^5$ & $5.1$ & $3.1$ & $3.6$ & $8.0$ & $12.0$ & 216 & 4 & 7.2 & 84 & 53 & 5 &142\\
3 & $3\times10^5$ & 6.0 & 3.6 & 6.4 & 13.7 & 20.1 & 251 & 5 & 6.8 & 164 & 94 & 15 & 273 \\
4 & $4\times10^5$ &      $6.2$   & $1.9$ & $11.4$ & $20.3$ & $31.7$ & 443 & 5 & 2.7 & 427 & 164 & 22 & 613 \\
\hline
$5^{\alpha}$ & $4\times10^5$ &  $5.8$ & $4.8$ & $9.8$ & $19.5$ & $29$ & 1016 & 3 & 6.1 & 156 & 73 & 13 & 242\\
$6^{\beta}$ & $2\times10^5$ & 1.2 & 12.9 & 0.7 & 5.0 & 5.7 & 107 & 4.6 & 55.7 & 10 & 118 & 21 & 149\\
\enddata
\tablecomments{Initial and final properties for all cluster models used in this study. Column 2 lists the initial number of planetary systems, $N_{\rm{P,\,i}}$, for each model. Columns 3 and 4 show the total number of single planets and planet--star binaries retained at 12 Gyr, respectively. Columns 5 and 6 show the cumulative number of planets ejected from their host cluster as isolated objects and as members of a planet--star binary. Column 7 shows the total number of planets ejected. Column 8 shows the number of distinct BH--planet binaries formed at late times (see Section \ref{sec:BHplanets}) and columns 9 and 10 show the median number of encounters and lifetimes of these BH--planet binaries, respectively. Columns 11, 12, and 13 show the number of BH--planet TDEs that occur through single--single (sin--sin), binary--single (bin--sin), and binary--binary (bin--bin) encounters, respectively and column 14 shows the total number of BH--planet TDE. For models 5 and 6, we vary cluster parameters aside from the initial number of planets. $^\alpha$For model 5, we assume an initial virial radius of 2 pc (as opposed to the fiducial value of $rv=1\,$pc used for other models). $^\beta$For model 6, we assume all primordial planet systems are born with an initial semi-major axis of 1 AU (as opposed to the fiducial value of 5 AU).}
\end{deluxetable*}

As planets (and stars) interact with BHs within the dense core of a cluster, they will occasionally undergo sufficiently close (gravitationally focused) encounters such that the planet or star may be tidally disrupted by the BH. In the standard picture, this requires a close passage in which the pericenter distance of the pair of objects is less than some characteristic tidal disruption radius, $R_{\rm{T}}$. The tidal disruption radius is related to the density of the object to be disrupted, and for objects with roughly uniform density such as low-mass stars and planets, can well-approximated using the classic formula of, e.g., \citet{Fabian1975}, $R_{\rm{T}} \approx (M_{\rm{BH}}/M)^{1/3} R$, where $R$ and $M$ are the radius and mass of the object to be disrupted. A number of recent analyses have explored the tidal disruption of \textit{stars} by BHs in dense star clusters and have shown these TDEs occur at a rate of up to $10^{-6}$ events per year in a Milky Way-like galaxy \citep{Perets2016,Lopez2018, Samsing2019, Kremer2019c}. Furthermore, it was shown in \citet{Kremer2019a} that in realistic clusters, BH--star TDEs most commonly involve M-dwarfs, simply because M-dwarfs are expected to dominate the mass function in typical GCs \citep{Kroupa2001}. Specifically, this analysis showed that the median mass of main-sequence stars disrupted by BHs is roughly $0.5\,M_{\odot}$.

Although Jupiter-like planets are comparable in radius to M-dwarfs, they are a factor of up to 100 times less massive. This implies the tidal disruption radius of Jupiter-like planets is roughly a factor of a few times larger than that of M-dwarfs. Thus, if the total number of planets and M-dwarfs is roughly comparable, we expect the rate of BH--planet TDEs to be larger than that of BH--star TDEs.

To quantify, the total rate of TDEs of planets by BHs through single--single encounters in a typical dense star cluster can be estimated by
\begin{multline}
    \label{eq:TDE}
    \Gamma_{\rm{TDE}} = \it{n}_{\rm{p}} (\pi R_{\rm{T}}^{\rm{2}}) v_{\infty} \Bigg[\rm{1}+\frac{\rm{2}\it{G} (M_{\rm{p}}+M_{\rm{BH}})}{\it{R}_{\rm{T}} v_{\infty}^{\rm{2}}} \Bigg] \it{N}_{\rm{BH}} \\
    \approx 40\,\rm{Gyr}^{-1} \Big( \frac{\it{n}_{\rm{p}}}{10^4\,\rm{pc}^{-3}} \Big) \Big(\frac{\it{R}_{\rm{p}}}{\it{R}_{\rm{J}}} \Big)  \Big(\frac{\it{M}_{\rm{p}}}{\it{M}_{\rm{J}}} \Big)^{-1/3} \\ \times  \Big( \frac{\it{M}_{\rm{BH}}}{20\,M_{\odot}} \Big)^{4/3} \Big( \frac{v_\infty}{10\,\rm{km/s}} \Big)^{-1} \Big( \frac{\it{N}_{\rm{BH}}}{1000} \Big).
\end{multline}
Here $R_{\rm{p}}$ and $M_{\rm{p}}$ are the radius and mass of a planet (we assume the mass and radius of Jupiter is typical). We have also assumed a typical BH of mass $20\,M_{\odot}$. Note that the characteristic values assumed for the number density of planets, $n_{\rm{p}}$, and the total number of BHs, $N_{\rm{BH}}$, within the mixing zone will be discussed further in Section \ref{sec:models} (specifically, see Figure \ref{fig:radial_dist}). Thus over the full lifetime of the cluster, we predict roughly 500 total BH--planet TDEs. 

In addition to occurring through single--single encounters, BH--planet TDEs can also occur during binary-mediated dynamical encounters (such as the planet--star binary plus BH single encounters described by Equation \ref{eq:BHstrong}), if during the encounter the BH and planet pass sufficiently close to one another. We explore the relative contribution of single--single encounters and binary-mediated encounters to the overall BH--planet TDE rate in Section \ref{sec:models}.

Figure \ref{fig:cartoon} illustrates the characteristic dynamical encounters relevant to the evolution of planet--star systems in GCs that have been discussed in this section.

\section{Globular cluster models}
\label{sec:models}

\subsection{Computational method}

To explore the interaction history of planetary systems in clusters, from initial distribution to final ejection or disruption, we create a set of cluster models using the cluster Monte Carlo code, \texttt{CMC} \citep{Joshi2000,Joshi2001, Fregeau2003, Umbreit2012, Pattabiraman2013, Chatterjee2010, Chatterjee2013a,Rodriguez2018}. Although the analytic predictions presented in the previous section provide motivation, a more detailed approach such as that provided by \texttt{CMC} is necessary to explore the intricacies of the dynamics relevant in a realistic cluster. \texttt{CMC} incorporates a variety of physical processes expected to play an important role in the large-scale structural evolution of the cluster as a whole as well as in the formation and evolution of BHs including two-body relaxation \citep{Henon1971a, Henon1971b}, stellar (and binary) evolution \citep{Hurley2000,Hurley2002,Chatterjee2010}, three-body binary formation \citep{Morscher2015}, galactic tides \citep{Chatterjee2010}, and small-$N$ gravitational encounters calculated using \texttt{Fewbody} \citep{Fregeau2004,Fregeau2007}, updated to incorporate post-Newtonian effects in all three- and four-body encounters \citep{Rodriguez2018}.

To treat the evolution of planetary systems, we follow the approach outlined in \citet{Chatterjee2012}. We assume that a fraction of low-mass ($M \leq \,M_{\odot}$) main-sequence stars are born with a Jupiter-like planet companion ($M_{\rm{p}}=M_{\rm{J}}$ and $R_{\rm{p}}=R_{\rm{J}}$). For simplicity, we assume all planet--star systems are born with a fixed semi-major axis (1 AU or 5 AU, depending on the model) and zero eccentricity. From a computational perspective, these planet--star systems are treated by \texttt{CMC} in the same manner as stellar binaries. For a detailed description of how binaries are evolved in \texttt{CMC}, see \citet{Fregeau2007}.

Note that we neglect here the complex processes relevant to the formation of planetary systems. These processes include a detailed treatment of the collapse of a protoplanetary disk, etc. which is well beyond the computational scope of \texttt{CMC}, however see, for example, \citet{Cai2019} for a discussion of some of these processes in the context of dense star clusters. Here we remain agnostic to the processes which govern the efficiency of planet formation at early times and simply vary the total number of planet--star systems at birth to explore the effect of the complex and interdependent dynamical processes inside a dense cluster. We also ignore tidal effects, which may occasionally lead to the formation of close binaries through tidal capture during close dynamical encounters \citep[e.g.,][]{Samsing2017,SamsingLeighTrani2018} and also eccentricity-migration and ensuing tidal interactions that may lead to the formation of close planet--star binaries, as explored in \citet{Hamers2017}. 

As discussed in Section \ref{sec:analytic}, when single planets and planet--star systems interact with stellar-mass BHs, a natural outcome is the tidal disruption of planets by the BHs. We handle BH--planet TDEs in the same manner as discussed in \citet{Kremer2019c} for BH--star TDEs. Here we assume in all models that dynamical encounters that lead to pericenter passages within the tidal disruption radius, $r_p \leq R_{\rm{T}} = (M_{\rm{BH}}/M_{\rm{p}})^{1/3} R_{\rm{p}}$, lead to a TDE.

We run a set of six independent cluster models. A variety of cluster parameters are fixed including the initial total number of objects, for which we assume $N=8 \times 10^5$, the initial King concentration parameter ($W_0 = 5$), metallicity ($Z=0.05Z_{\odot}$), galactocentric distance ($r_g=8\,$kpc), and stellar binary fraction ($f_b=5\%$). As shown in several previous analyses \citep[e.g.,][]{Kremer2019a}, these various cluster properties result in GCs that are similar to typical Galactic GCs at the present-day. We vary the total number of planet--star systems at birth ($N_p=10^5, 2\times 10^5, 3\times10^5,$ and $4\times 10^5$), the initial cluster virial radius ($r_v=1$ and 2 pc), as well as the orbital separation of planetary systems at birth ($a=1\,$AU and 5 AU). Table \ref{table:models} lists the initial parameters of all models as well as various model properties at 12 Gyr.

\subsection{Planet populations throughout cluster lifetime}

\begin{figure}
\begin{center}
\includegraphics[width=0.5\textwidth]{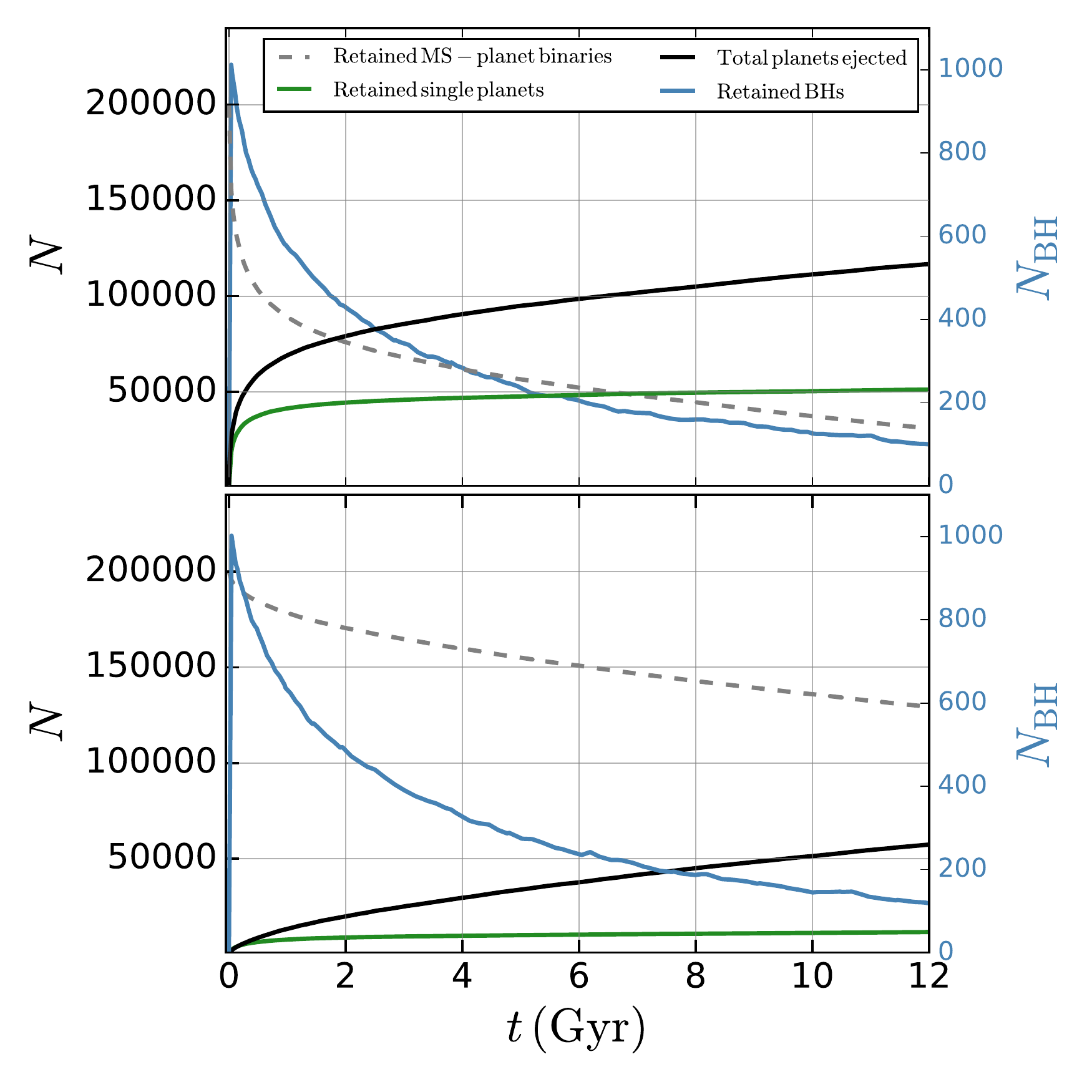}
\caption{\label{fig:cluster_properties} \textit{Top panel:} The time evolution of the number of retained MS--planet binaries (dashed gray), retained single planets (which were unbound from their primordial MS companions through dynamical encounters with other stars; green), cumulative number of planets ejected (black), and total number of retained BHs (blue) for model 2 in Table \ref{table:models}. \textit{Bottom panel:} Same as top panel but for model 6. Models 2 and 6 are identical except for the orbital separation assumed for planetary systems at birth ($a=5\,$AU and 1 AU, respectively).}
\end{center}
\end{figure}

\begin{figure*}
\begin{center}
\includegraphics[width=\linewidth]{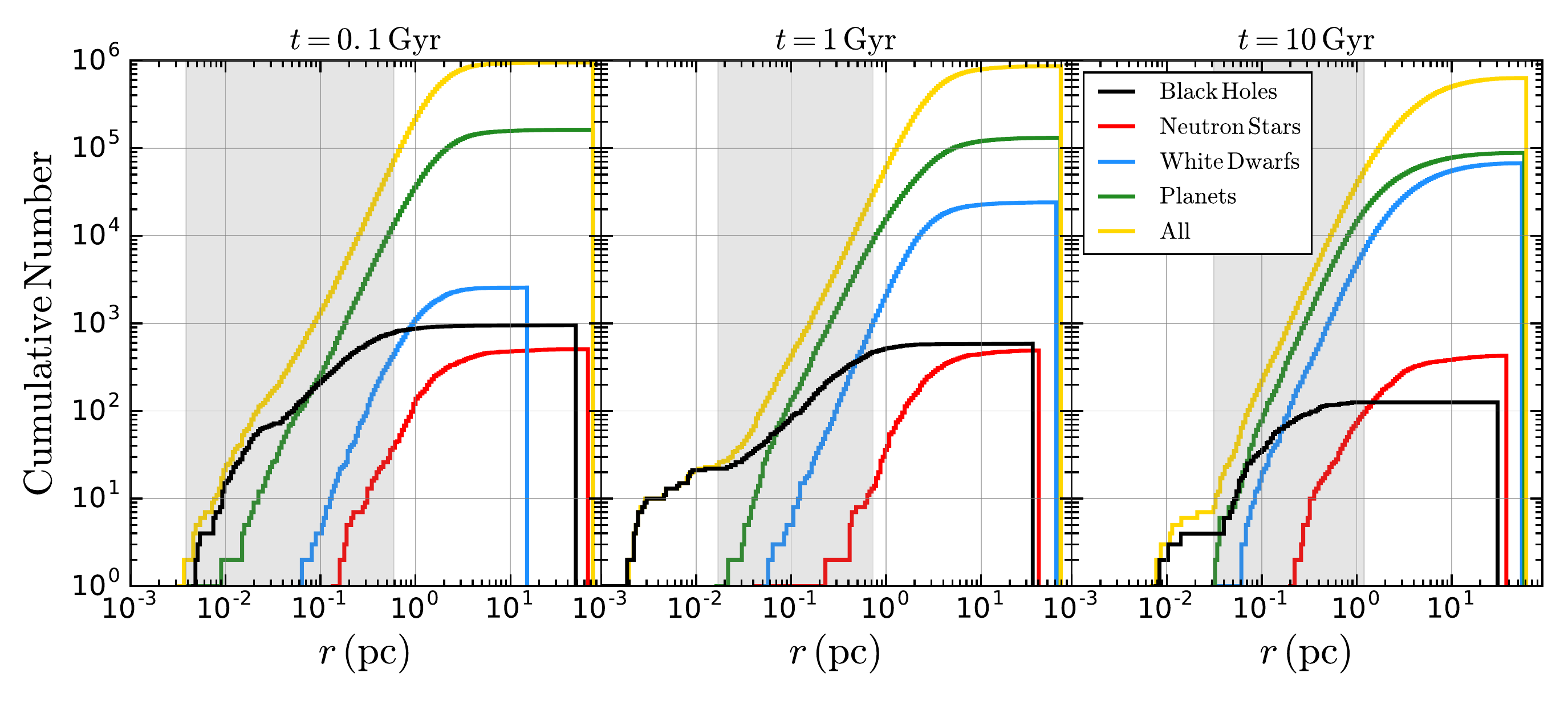}
\caption{\label{fig:radial_dist} Radial profiles of planets (green), BHs (black), neutron stars (red), white dwarfs (blue), and all stars (yellow) for model 2 at three different time snapshots (from left to right: 100 Myr, 1 Gyr, and 10 Gyr). The shaded gray regions give a visual representation of the mixing zone where the BH and planet populations overlap, as described in the text. Note that the BH and planet populations overlap radially throughout the entire evolution of the cluster, setting the stage for dynamical formation of BH--planet binaries and BH--planet TDEs.}
\end{center}
\end{figure*}

 In Figure \ref{fig:cluster_properties}, we show the time evolution of various cluster populations for model 2 (top panel) and model 6 (bottom panel) in Table \ref{table:models}. In dashed gray, we show the number of MS--planet binaries retained in the cluster, in green, the number of retained single planets, in black, the cumulative number of planets that have been ejected from the cluster, and in blue (the right-hand vertical axes), we show the total number of retained BHs in each model. The only difference between the two models shown in Figure \ref{fig:cluster_properties} is the orbital separation assumed for the planetary systems at birth (5 AU for model 2 and 1 AU for model 6).

From Equation \ref{eq:strong}, we see that the rate of dynamical encounters increases with the semi-major axis. Therefore, we expect that by adopting a smaller value for the orbital separation, fewer planetary systems will undergo encounters that can potentially unbind the planets from the primordial companion star. Figure \ref{fig:cluster_properties} reveals this to be the case: inspection of the figure shows that about a factor of 5 more planet--star binaries and roughly a factor of 5 less single planets are found in the $1\,$AU case compared to the $5\,$AU case, as expected from the factor of 5 change in semi-major axis between the two models.

Comparing the top and bottom panels, we see that $1.3\times10^5$ planet--star binaries (roughly 65\% of the initial population) remain intact by 12 Gyr for the $a=1\,$AU assumption, versus only $3\times10^4$ planet systems (roughly 15\% of the initial population) for the $a=5\,$AU assumption. Likewise, a much small number of planets have been ejected from the cluster ($5.7\times10^4$ versus $1.2\times10^5$) when a smaller initial orbital separation is assumed. As can be seen from inspection of columns 5 and 6 of Table \ref{table:models}, of the planets that are ejected from the cluster in model 6, a relatively high fraction are ejected as a member of a binary (i.e., they are still bound to their host star) as opposed to being ejected as single planets that have been unbound from their host stars. This is anticipated: more compact planet systems have higher binding energies and thus are less likely to be unbound from their host star in the event of a dynamical encounters.  

Planet--star binaries can escape from their host cluster through two mechanisms: dynamical ejections, where the binary acquires sufficient energy from a dynamical encounter to escape the potential well of the cluster, and tidal loss through mass-segregation, where the object's radial position simply extends beyond the cluster's tidal radius. Although we allow for both methods in \texttt{CMC} (see \citet{Chatterjee2010} for review of our tidal treatment), for simplicity, we do not differentiate the systems that escape through these two mechanisms in Figure \ref{fig:cluster_properties}. We do note here that for models 1--5 (which adopt $a=5\,$ AU for initial planet orbits) roughly 60\% (40\%) of planetary systems escape through dynamical ejections (tidal loss) while for model 6 (which adopts $a=1\,$AU), roughly 20\% (80\%) escape through dynamical ejections (tidal loss). As low-mass objects, all planet--star binaries naturally tend toward the outer parts of their host cluster through mass-segregation processes where escape through the tidal boundary becomes more likely. However, wider planet--star binaries are more likely to undergo strong dynamical encounters en route to the tidal boundary which contributes to larger numbers of dynamically-ejected planets. We reserve for a future analysis a more careful study of the specific mechanisms through which planets are ejected from clusters and, in particular, how the relative rates of these mechanisms change with initial planet properties and cluster properties.

The final key point illustrated in Figure \ref{fig:cluster_properties} is that the total number of BHs decreases throughout the lifetime of the cluster. Due to recoil kicks attained through series of strong dynamical interactions with one another in a cluster's core, BHs will naturally be ejected from their host cluster, slowly depleting the overall BH population. See, for example, \citet{Morscher2015} for further detail.

In Figure \ref{fig:radial_dist}, we show the radial distributions of several populations in model 2 at three snapshots in time: 0.1 Gyr, 1 Gyr, and 10 Gyr. In yellow, we show the radial distribution of all objects in the cluster, in green we show the distribution of planets (both single and bound planets combined), in black we show BHs, in blue we show white dwarfs (WDs), and in red we show neutron stars (NSs).

As also illustrated in Figure \ref{fig:cluster_properties}, Figure \ref{fig:radial_dist} shows that the total number of BHs and total number of planets decreases throughout the evolution of the cluster through dynamical processes discussed earlier. This is in contrast to, for example, the WD population which is seen to increase throughout the lifetime of the cluster, simply because of stellar evolution of low-mass stars.

As shown in all three panels, the inner part of the cluster ($r \lesssim\rm{a\,few} \times 10^{-2}\,$pc) is dominated at all times by BHs. These BH-cores, which are a simple consequence of mass-segregation, have been well-studied in a variety of recent analyses \citep[e.g.,][]{Morscher2015,Askar2018,ArcaSedda2018}. 

\begin{figure}
\begin{center}
\includegraphics[width=0.47\textwidth]{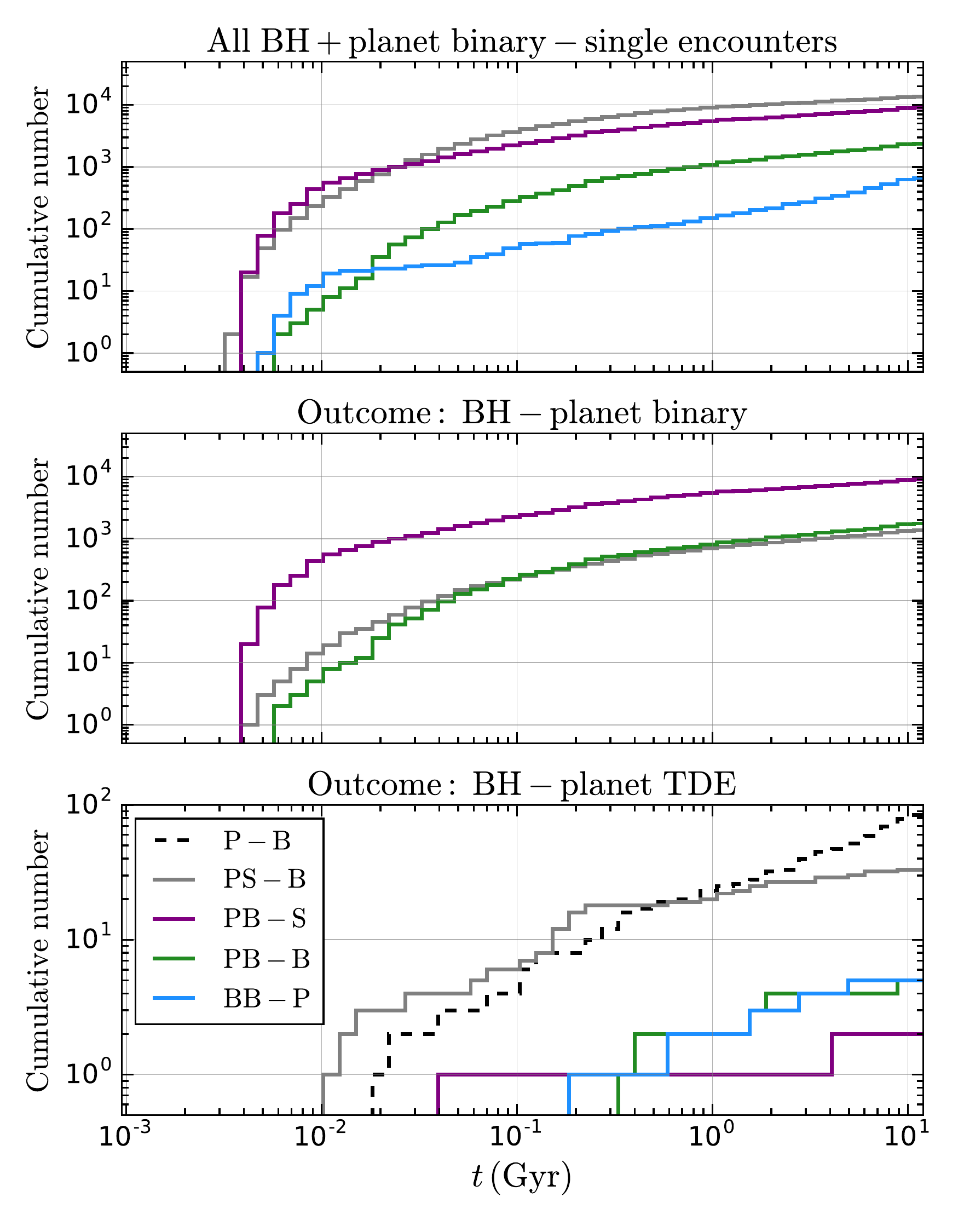}
\caption{\label{fig:binsin} \textit{Top panel:} Cumulative distribution of various types of binary--single encounters involving planets and at least one BH. Here ``P'' denotes planet, ``B'' denotes BH, and ``S'' denotes a star (non-BH). \textit{Middle panel:} Binary--single encounters of various types that lead specifically to a stable BH--planet binary outcome. \textit{Bottom panel:} Encounters that lead specifically to a BH--planet TDE.}
\end{center}
\end{figure}

The gray shaded regions in Figure \ref{fig:radial_dist}, denote what we define as the ``mixing zone'' of BHs and planets within the cluster. For illustrative purposes, we define this mixing zone as a radial shell whose inner radius is determined by the radial position of the innermost planet and whose outer radius is determined by the radial position of the cluster’s outermost BH that lies within the observed core radius of the cluster,\footnote{As a result of recoil from scattering encounters, some BHs can be found outside the cluster's core radius before they sink back to the core through mass segregation. We ignore these outermost BHs because, the interaction cross section is expected to be dominated by only the BHs within the core where the stellar density is relatively high.} following the definition in \citet{Kremer2018a}. As all three panels demonstrate, the BH and planet populations overlap with one another throughout the entire lifetime of the cluster, leading to BH--planet dynamical interactions. These interactions can lead to both the formation of BH--planetary systems (through exchange encounters) and BH--planet TDEs. 

To demonstrate the types and relative rates of various BH--planet encounters that may occur in a typical cluster, we show in Figure \ref{fig:binsin} the various types of binary--single encounters\footnote{As mentioned previously, binary--binary encounters also play a role in these various channels. We focus on binary--single here for simplicity but again emphasize that strong binary--binary encounters are also integrated within our \texttt{CMC} models.} involving planets and at least one BH that occur in model 2. In the top panel, we show the cumulative time distribution of all such encounters, independent of outcome. In gray we show planet--star binary plus BH single (``PS--B''), in purple we show planet--BH binary plus star (non-BH) single (``PB--S''), in green we show planet--BH binary plus BH single (``PB--B''), and in blue we show BH--BH binary plus planet single (``BB--P''). In the middle and bottom panels of the Figure \ref{fig:binsin}, we show the distribution of these binary--single encounters that lead to the specific outcomes of a stable BH--planet binary and a BH--planet TDE, respectively. In the bottom panel, we also show as a dashed black curve the distribution of BH--planet TDEs that occur through single--single encounters. We discuss BH--planet TDEs in more detail in Section \ref{sec:EM}.

As shown by the gray curve of the top panel, PS--B encounters are the first to occur. These encounters can lead to formation of planet--BH binaries through exchange and the specific number of these encounters that lead to planet--BH binaries can be seen from comparisons of the gray curves of middle and top panels: for this model, we find 1365 out of roughly 13500 (roughly 10\%) PS--B binary--single encounters result in the formation of a planet--BH binary. Once formed, planet--BH binaries can undergo repeated encounters (on average 5; see column 9 of Table \ref{table:models}) before being broken. In the middle panel we do not distinguish between binary--single encounters that lead to the formation of a \textit{new} planet--BH binary versus encounters that preserve a planet--BH that entered the encounter. Thus, because the majority of PB--B and PB--S encounters result in preservation of the original binary, the PB--B and PB--S curves in the top and middle panels are nearly identical.

Once unbound from their host star, single planets can undergo encounters with BH--BH binaries (labeled BB--P encounters in Figure \ref{fig:binsin}). Because the planet is much less massive than either BH component, that probability that the planet will exchange into a binary with one of the BH components is vanishingly small. However, unlike all other BH--planet encounters considered here, BB--P encounters can have total negative energy which means a resonant encounter is possible. This is simply because the internal potential energy of a BH--BH binary is typically much larger in magnitude than that kinetic energy of an incoming single planet. In this case, the planet can undergo multiple passages before ultimately being ejected from the three-body system. We discuss this specific encounter further in Section \ref{sec:BBH-TDEs}.

\subsection{Dynamical formation of BH--planetary systems}
\label{sec:BHplanets}

\begin{figure}
\begin{center}
\includegraphics[width=0.5\textwidth]{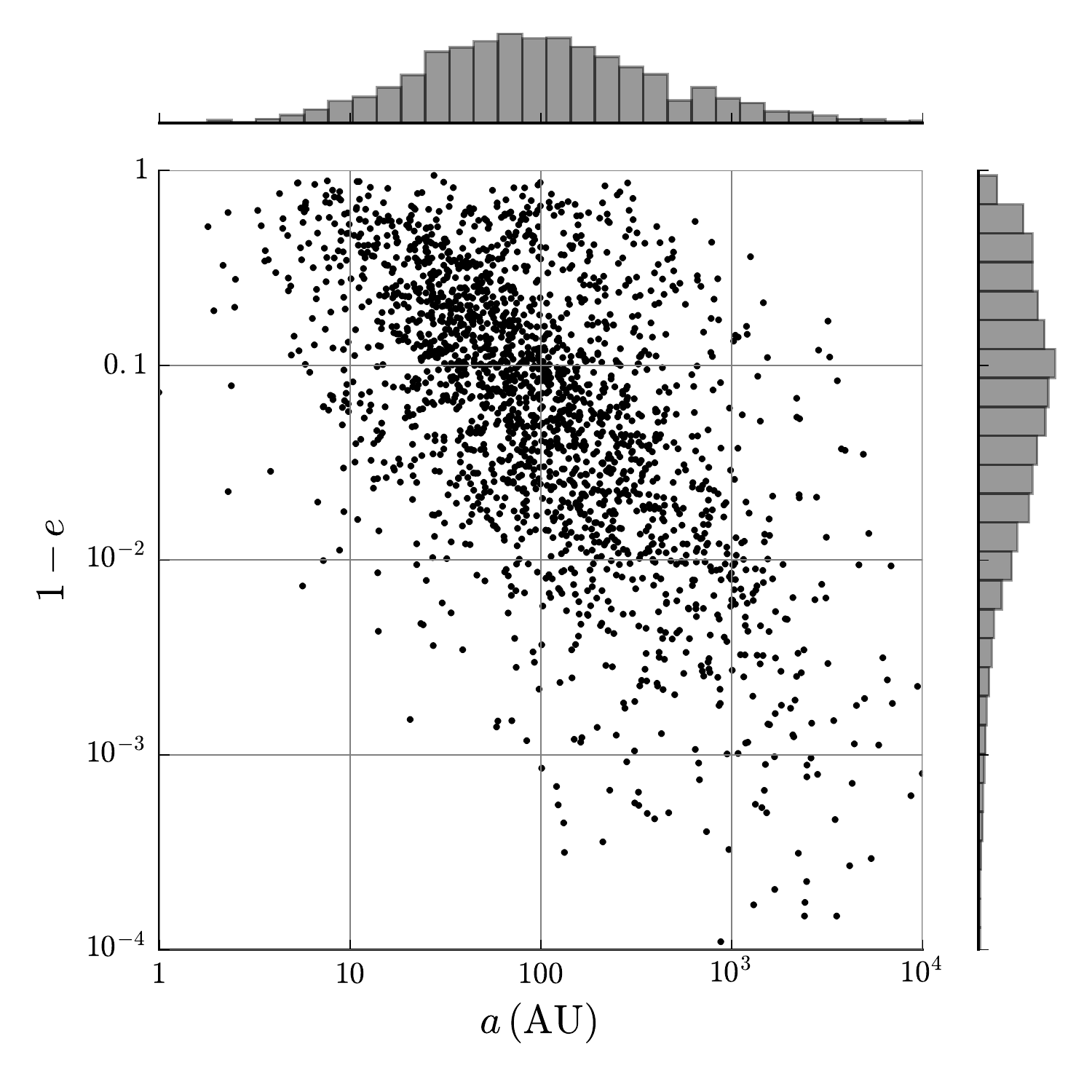}
\caption{\label{fig:BHplanets} Orbital parameters for all distinct BH--planet binaries formed through binary interactions at late times (8-12 Gyr; chosen to reflect the uncertainty in present-day ages of Milky Way GCs).}
\end{center}
\end{figure}

For old GCs, we are specifically interested in those BH--planet binaries that may reside in the clusters at late times; BH--planet binaries formed early in the evolution of the cluster are unlikely to survive to the present. Column 10 of Table \ref{table:models} shows the total number of distinct BH--planet binaries that form at late times in our five cluster models through the various binary--single encounters described in the previous subsection and also through binary--binary encounters. We define ``late times'' as 8-12 Gyr, which reflects roughly the uncertainty in ages of GCs in the Milky Way. Figure \ref{fig:BHplanets} shows the semi-major axis and eccentricity for all such binaries, as seen in the models.

In total, roughly 2000 distinct BH--planet binaries are identified in our models. As seen in Table \ref{table:models}, the number of BH--planets increases with the total number of planets at birth, as expected. Once they have been dynamically formed, the mean lifetime of these BH--planet binaries (before they are broken apart by subsequent encounters) ranges from roughly $3-50\,$Myr (depending on the various model parameters; see column 10 of Table \ref{table:models}). Thus at any given snapshot in time, the total number of BH--planet binaries present is unlikely to be more than a few. This is consistent with our understanding of the rates at which BH-non-BH binaries will dynamically form in GCs, as discussed in e.g., \citet{Kremer2018a}.

As also shown in Figure \ref{fig:radial_dist}, the planet population also overlaps with WDs and NSs throughout the cluster's lifetime. This allows planets to form binaries with both NSs and WDs through similar mechanisms as discussed here for BHs. We briefly discuss the implications of mixing of planets with WDs and NSs in Section \ref{sec:NSplanets}.

\section{Black hole--planet TDE\lowercase{s}}
\label{sec:EM}

In this section we discuss the number of BH--planet TDEs identified in our models and estimate the expected rate of such events in the local universe (Section \ref{sec:rates}). We go on in Section \ref{sec:detection} to calculate the expected electromagnetic signature associated with these events and estimate the prospects for detection by upcoming all-sky surveys such as LSST. We finish in Section \ref{sec:BBH-TDEs} with a discussion of the special case where a planet TDE occurs during a resonant encounter with a binary BH.

\subsection{Rates}
\label{sec:rates}

Columns 11-14 in Table \ref{table:models} show the total number of BH--planet TDEs identified in each cluster model through both single--single encounters and binary-mediated interactions (refer also to Figure \ref{fig:binsin}). As anticipated, the number of BH--planet TDEs increases as the number of planets increases. However, as seen from a comparison of models 4 and 5, the number of BH--planet TDEs also depends upon the cluster's initial size. More compact clusters have higher densities and thus a higher rate of dynamical encounters.

We find as many as roughly 600 BH--planet TDEs can occur throughout the lifetime of a single cluster. Scaling to the Milky Way GC population, this corresponds to a BH--planet TDE rate of roughly $10^{-5}\,\rm{yr}^{-1}$ for a Milky Way-like galaxy, a factor of roughly a few times higher than the BH--star TDE rate predicted in \citet{Kremer2019c}. 

The fraction of stars that host planetary companions at birth is highly uncertain. To reflect this uncertainty we vary this fraction in our models (column 2 of Table \ref{table:models}). As this is a first study, we assume a planet-to-star ratio, $F_{\rm{p}}$, of at most 0.5, primarily to ease computational cost. But it is very possible (and perhaps even realistic) that the primordial planet-to-star ratio may be as high as one or more. Previous work \citep[e.g.,][]{HurleyShara2002}, has even suggested planets may outnumbers stars by a factor of up to 100. In this case, the rate of BH--planet TDEs may be significantly higher than that predicted from the models in this study. 

For now, to estimate how the TDE rate may increase with larger values of $F_{\rm{p}}$, we simply fit the $N_{\rm{p}}-N_{\rm{TDE}}$ relation in Table \ref{table:models} to a power law and extrapolate to higher values of $F_{\rm{p}}$. For a fixed number of stars, we find the relation $N_{\rm{TDE}} \approx 1500\it{F}_{\rm{p}}^{\rm{1.6}}$ gives a rough estimate of the scaling. Using this relation, we predict that for $F_{\rm{p}}=1$, the BH--planet TDEs rate may be as high as roughly 1500 events over the lifetime of a single cluster and up to roughly a few$\times10^{-5}$ events per year in a Milky Way-like galaxy. Detailed consideration of the inclusion of larger numbers of planets as well as consideration of the specific scaling of the TDE rate with planet fraction (and other cluster parameters such as cluster mass), is beyond the scope of this study and we hope to explore some of these ideas further in a later paper.

\begin{figure}
\begin{center}
\includegraphics[width=0.5\textwidth]{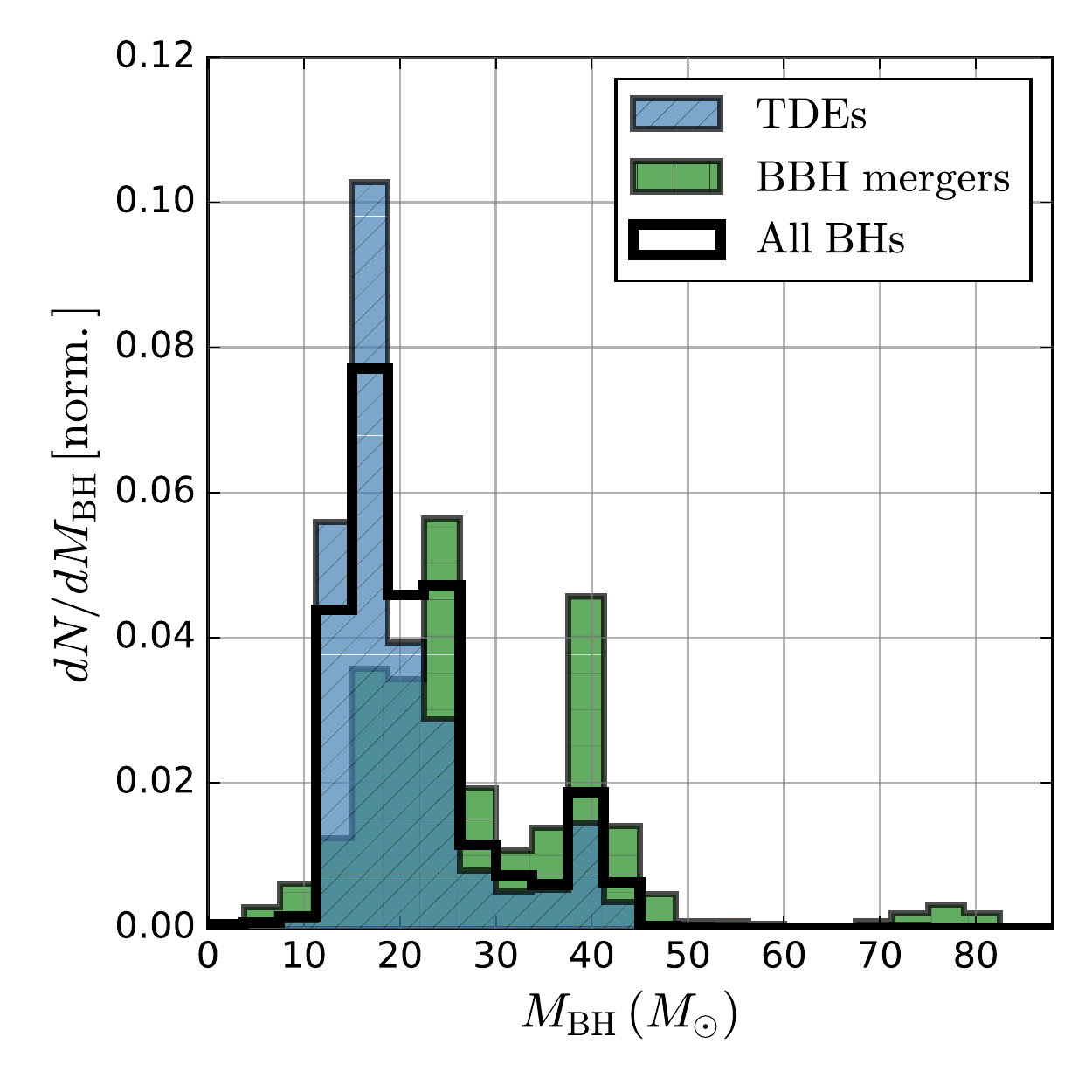}
\caption{\label{fig:BH_masses} Mass spectrum for various BH populations. The black histogram shows masses of all BHs retained in the cluster initially, green shows masses of all BHs that eventually undergo BBH mergers (including second-generation BHs), and blue shows masses of all BHs that undergo a BH--planet TDE.}
\end{center}
\end{figure}

Figure \ref{fig:BH_masses} shows the mass distributions for all BHs that undergo BH--planet TDEs in our models (blue histogram) compared to the mass distribution for all BHs retained upon formation in the clusters (black), and the mass distribution for all BHs that eventually undergo binary BH (BBH) mergers. The prominent peak at $M_{\rm{BH}}\approx40\,M_{\odot}$ comes from our treatment of (pulsational) pair instability supernovae \citep[see][for details]{Rodriguez2018a}. The small peak in the BBH-merger mass distribution at $M_{\rm{BH}}\approx 80\,M_{\odot}$ comes from so-called ``second generation'' mergers \citep[e.g.,][]{Rodriguez2019}. As the figure shows, on average, the most massive BHs preferentially undergo BBH mergers, while the lower mass BHs preferentially undergo TDEs. This is anticipated: the most massive BHs efficiently mass-segregate upon formation and are driven to merger first (see \citet{Rodriguez2018b,Samsing2018a} for a summary of the various channels for BBH mergers in typical GCs). Meanwhile, the least massive BHs are more likely to mix with other stellar populations and thus facilitate BH--planet TDEs (see also Figure \ref{fig:radial_dist}). This is analogous to the result in \citet{Kremer2018a} in the context of forming accreting BH binaries. Overall, the average mass of merging BHs is roughly $30\,M_{\odot}$ while the average mass of BHs that facilitate TDEs is $20\,M_{\odot}$.

\subsection{Prospects for detection}
\label{sec:detection}

The fundamental assumptions that lead to the prediction of the fallback timescale and luminosity of a tidal disruption of a main-sequence star by a supermassive BH are valid in the case of a stellar mass BH disrupting a giant planet, as was pointed out by \citet{Perets2016}.

While the size of the planet is much larger than the event horizon of the disrupting BH in the planet case,
\begin{equation}
 \frac{R_J}{R_S} \approx 1183 \left(\frac{M_{\rm{BH}}}{20\Msun} \right)^{-1},
\end{equation}
the more relevant ratio, between the planet radius and the disrupting pericenter passage radius $r_p$, is still in the regime where the planet is small compared to the size of the disrupting orbit. That is,
\begin{equation}
 \frac{R_J}{r_p} = \beta \frac{R_J}{R_T} \approx 0.037 \beta \left( \frac{ q }{ 3.33 \times 10^{-5} }\right)^{1/3} 
\end{equation}
for small values of the penetration factor $\beta = R_T/r_p$, where $q\equiv M_J/M_{\rm{BH}}$ and $R_T = q^{-1/3} R_J$ is the tidal radius. For a disruption, $\beta\geq1$. 

Hence, for encounters where $\beta \lsim 30$, the disrupted material will stay in a ballistic orbit until returning to pericenter, where it shocks and generates a tidal disruption flare.
For $\beta \gsim 30$, the giant planet will partially envelope the BH at pericenter. This could possibly lead to an intense accretion event that may even outshine the corresponding tidal disruption flare caused by fallback.

Measuring $\beta$ for each of the BH--planet disruptions in the \texttt{CMC} simulations, we find that $\beta\lsim30$ ($\lsim 10$) for $\sim 95 \%$ ($\sim 90 \%$) of all disruptions. Hence, we leave investigation of the close-encounter planet disruptions for future work and now focus on the fallback flare. 

The tidal disruption of a giant planet by a stellar mass BH also occurs at a larger number of gravitational radii ($r_G$) from the BH than in the supermassive BH--star case. For the fiducial planet disruption considered above, compared to the disruption of a Sun-like star by a supermassive BH with mass $M_{\rm SBH}$, 
\begin{equation}
\frac{ \left[R_T/r_G\right]_{\rm plt+sBH} } { \left[R_T/r_G\right]_{\rm str+SBH} } \sim 1364 \left( \frac{M_{\rm SBH}}{10^6 \Msun} \right)^{2/3}
\end{equation}
meaning that strong gravity effects such as orbital precession are much weaker in this case. 
This could have implications for stream circularizion and the evolution of the debris compared to the supermassive BH case \citep[e.g.,][]{HayasakiStone+2016}.

Another possible difference from the star--supermassive BH treatment is the orbital energy at infinity for these systems, which are disrupting via flybys and (less commonly) three body interactions. The standard treatment assumes parabolic orbits, with zero energy at infinity, resulting in a $t^{-5/3}$ fallback rate and half of the debris remaining bound.

We argue that, here too, the parabolic orbit is a good approximation. This is because the majority of disruptions occur for non-resonant interactions (Figure \ref{fig:binsin}), in which case we can estimate a hyperbolic eccentricity of the disrupting orbit by equating the velocity dispersion of BH-planet populations with the hyperbolic excess velocity. In this case, the typical hyperbolic eccentricity of a disrupting encounter is small,
\begin{multline}
e_{\rm hyp} - 1 \approx \frac{GM_{\rm BH}}{v^2_\infty} \beta R_T \\ \approx 7 \times 10^{-5} 
\beta
\left(\frac{M_{\rm BH}}{2 0\Msun} \right)\left(\frac{v_\infty}{10 \rm{km/s}} \right)^{-2},
\end{multline}
where we adopt $\beta=1$ throughout.

Comparing this to a critical hyperbolic eccentricity, above which none of the planet will be bound to the BH after disruption \citep{Hayasaki+2018},
\begin{equation}
    e_{\rm crit} - 1 \approx  \frac{2q^{1/3}}{\beta} \approx 0.06 \beta^{-1},
\end{equation}
we see that in the case at hand, and for $v_{\infty} \lesssim 300$km/s, parabolic orbits are sufficient for modelling disruptions from flyby-like encounters.

Despite small differences, the above reasoning suggests that, at least for the most common single-BH--planet TDEs, we may estimate the duration and peak luminosity of the fallback flare as in the standard literature for stars disrupted by supermassive BHs \citep{Rees1988, Phinney:1989}.

Once a planet of mass $M_{\rm p}$ and radius $R_{\rm p}$ is disrupted, the time for orbiting material to return to the disruption point provides the approximate time scale for the rise of the flare, and hence a characteristic timescale for the emission,
\begin{equation}
\tau \approx 1.1 \ \rm{days} \left( \frac{M_{\rm p}}{M_J} \right)^{-1} \left( \frac{R_{\rm p}}{R_J} \right)^{3/2} \left( \frac{M_{\rm{BH}}}{20\Msun} \right)^{1/2},
\end{equation}
so that a typical flare will take place over the course of a few days (2 days to drop to half of the peak luminosity, 5 days to drop an order-of-magnitude in luminosity).

The peak luminosity, at time $\tau$ after disruption, is
\begin{equation}
\mathcal{L} \approx 2.4 \times 10^{40} \rm{erg}\ \rm{s}^{-1} \left( \frac{M_{\rm p}}{M_J} \right)^{7/3} \left( \frac{R_{\rm p}}{R_J} \right)^{-5/2} \left( \frac{M_{\rm{BH}}}{20 \Msun} \right)^{1/6}.
\label{Eq:Lpeak}
\end{equation}
where we have followed \citep{LiNarayanMenou:2002} and estimated the radiation efficiency as the specific energy of the circularizing orbit at $2 R_T$. Compared to the case of a supermassive BH and star, where radiation efficiencies are a few percent, the planet--stellar-mass BH case has much lower efficiencies of $\sim 10^{-6}$. Again we have assumed that half of the planet stays bound after disruption; the peak luminosity scales linearly with this fraction.

\subsubsection{Observation with upcoming all sky-surveys}

For a rate of $10^{-5}\,\rm{yr}^{-1}$ per Milky Way-like galaxy, the planet disruption rate is only an order of magnitude lower than the rate of stellar TDEs in galactic nuclei of $10^{-4}\,\rm{yr}^{-1}$ per galaxy \citep[e.g.,][]{vV18}, and may rival this rate for planet fractions larger than the conservative values simulated here. However, the peak luminosity is $\sim10^3$--$10^5$ times lower. 

Assuming $10^{-2}$ Milky-Way-like galaxies per Mpc$^{-3}$ \citep{MDPRada:2009}, detection becomes probable only if the entire sky can be surveyed at a few days cadence out to $\sim 150$~Mpc.

LSST will cover a 18000 deg$^2$ at a cadence of roughly once per 4.5 days at a limiting magnitude of $\sim 24.5$ \citep{LSST:2019}.
Using the bluest (high-efficiency) filter from LSST (g-band) we compute a maximum distance to which LSST can detect TDEs with an assumed isotropic peak luminosity predicted in Eq. (\ref{Eq:Lpeak}),
\begin{eqnarray}
d_{\rm LSST} &\lsim & 180 \rm{Mpc} \left(\frac{\mathcal{L}}{2.4 \times 10^{40} \rm{erg}\ s^{-1}} \right)^{1/2}
\left(\frac{ F_{\rm min} }{ F_{g, \min} } \right)^{1/2} \nonumber \\
F_{g, \min} &\approx& \nu_{g} F_{g,0} 10^{-m_{g,\min}/2.5}
\end{eqnarray}
where $\nu_g \approx 6.338 \times 10^{14}$~Hz and $F_{\rm{g},0} \approx 3.92\times 10^{-20}$ erg s$^{-1}$ cm$^{-2}$ Hz$^{-1}$ \citep{LSSTSVO1, LSSTSVO2}, and we use a limiting LSST magnitude of $m_{g,\min}=24$. We note however that deep LSST observations can reach magnitudes of 27 (increasing our maximum distance to a Gpc), but at lower cadence and covering a smaller fraction of the sky. 
Assuming that all of the luminosity comes out in the g-band, our estimate of the peak luminosity implies that, at best, LSST could detect $\sim 2$ planets from GCs per year, amounting to 10's of detections over its lifetime.

As discussed in Section \ref{sec:rates}, the primordial planet-to-star ratio assumed in our models may underestimate the total number of planets in young massive clusters. If the planet-to-star ratio is increased from our maximum value of 0.5 to a value of 1, the detection rate of BH--planet TDEs may increase by a factor of a few and up to roughly 10 TDEs per year may be realistic for LSST. If in addition to primordial planetary systems, a young cluster also has a significant population of single planets \citep[formed through, for example, planet--planet interactions;][]{Rasio1996}, this rate may increase even further. In such a case, the Zwicky Transient Facility \citep[ZTF;][]{ZTF:2014}, which will reach a limiting magnitude of $m \lesssim 21$ with up to a few days cadence, may detect a few BH--planet TDEs in GCs during its tenure. We reserve more detailed consideration of these possibilities for a future study.

\subsubsection{Identification of BH--planet TDEs}

As with stellar TDEs, a main difficulty in detection is identification of a flare with a TDE, and not some other transient. Additionally, we would like to differentiate planetary and stellar TDEs. Here we briefly discuss a few ways that BH--planet TDEs could be identified as such.

Our simple estimation of TDE observables predicts the standard TDE flare and power law decay. 
Because the BH mass and mass of the disrupted object enter into the fallback rate, parameter estimation from the observed light curves within the fallback model would give small values for the disrupting BH and disrupted object mass \citep{Mockler+2019}. That is, the characteristic timescales and luminosity of disruption in the planet--BH case are smaller than for the scaled up star--supermassive BH case. In addition, such disruptions are not confined to occur in galactic nuclei (though they may if similar planet--BH interactions occur in nuclear-star clusters); rather, they would most probably occur in halos of galaxies where the globular clusters reside (requiring $\sim 10$ arcsecond resolution at $200$~Mpc to determine).

While we do not go into detail in this work, the spectrum of a disrupted gas giant may differ from the stellar case if the planet core is large and contains metals. Lithium, for example, which is easily destroyed in stars, could play a larger role in the spectra of disrupted planets, which will not have reduced their Lithium abundances over time \citep[e.g.,][]{BasriLi:1998}.

In order to further set apart planet TDEs from other day-long transients and disruptions of stars, future work should consider the effects of different equations of states \citep[as in][]{Lodato:2009} and a large planet core on the disruption outcome. Additionally, the relatively weak field gravity resulting in lack of orbital precession should be further investigated in the planet case as generating unique identifiers of a planet TDE \citep[one such TDE was simulated in][]{Perets2016}.

Another interesting possibility for discerning a planet TDE from a stellar TDE could be variability in the decaying light curve from disruption by a stellar-mass BBH (which make up $5\%$ of disruptions identified in this models of this study). While much less likely, one could imagine a BBH merger accompanied by a planet resulting in an electromagnetic counterpart disruption. We discuss BBH disrupters further in the next subsection.

\subsection{Tidal disruption by binary BHs}
\label{sec:BBH-TDEs}

\begin{figure}
\begin{center}
\includegraphics[width=0.5\textwidth]{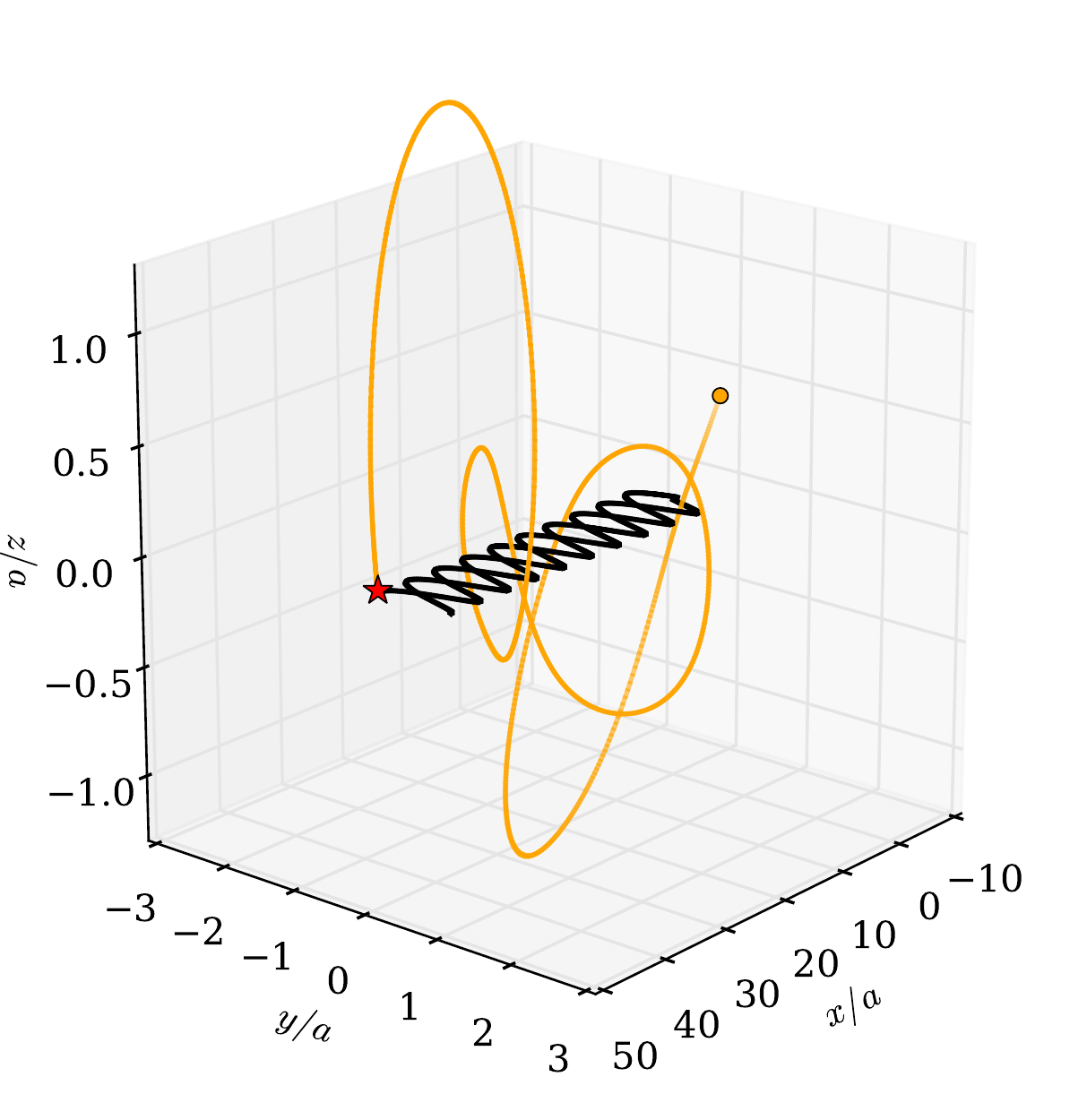}
\caption{\label{fig:BBH-planet} Example of a resonant binary--single encounter with a BBH and incoming single planet. For the initial BBH, we assume $a=0.5\,$ AU, $e=0$, and masses $M_1=M_2=30\,M_{\odot}$. The red star denotes the moment that the planet is tidally disrupted by a close passage to one of the BHs.}
\end{center}
\end{figure}

As discussed in Section \ref{sec:dyn_disruption}, the strong binary--single encounters involving BHs and planet--star binaries rarely occur in the resonant regime, where $v_\infty < v_c$, simply because $v_c$ for a binary with one planet component is low ($\lesssim 1\,\rm{km\,s}^{-1}$). However, in the special case where the incoming star is a single planet and the binary is a compact BBH, $v_\infty$ can easily be much less than $v_c$, and a resonant encounter is possible. In this case, the planet may undergo multiple passages of the BBH before becoming unbound or passing sufficiently close to one of the BHs to become tidally disrupted. If a TDE occurs in this scenario (which we refer to as a BBH--planet TDE), the TDE may impact the subsequent evolution of the BBH. This is qualitatively similar to the scenario described in \citet{Lopez2018, Samsing2019} for TDEs of stars by
BBHs in GCs.

In total, we identify 55 BBH--planet TDEs in our models, which constitute roughly $13\%$ of all planet TDEs occurring through binary--single encounters and $5\%$ of all BH--planet TDEs in general. In Figure \ref{fig:BBH-planet}, we show an example binary--single encounter which leads to a BBH--planet TDE. Here, we assume a BBH with initial $a=0.5$ AU, $e=0$, and masses $M_1=M_2=30\,M_{\odot}$.
As seen, the incoming planet becomes temporarily bound to the BBH, which leads to multiple close encounters. In the majority of cases, the interaction concludes with the planet getting ejected through a slingshot interaction by one of the BHs; however, in this case shown in the figure, the planet is disrupted (shown with a {\it red star}) before ejection. As described in \citet{2017ApJ...846...36S, Samsing2019}, the cross section for disruption is, to leading order, independent of the semi-major axis of the disrupting BBH, $a$, and as a result, is simply proportional to $M_{\rm{BH}} R_{\rm T}$. Therefore, the relative rate of BBH--planet TDEs does not depend strongly on the BBH orbital distribution. However, the outcome of individual TDEs and their corresponding observables do depend on the orbital parameters of the BBH \citep[e.g.][]{Lopez2018, Samsing2019}. For example,
the relative energy between the disrupting BH and the planet increases as $a$ decreases due to the corresponding increase in the orbital velocity of the disrupting BH.

The fact that the interaction of planets and stars with the BBH population occasionally lead to TDEs,
makes it possible to indirectly probe properties of the BBH population, if the corresponding electromagnetic (EM)
signal can be detected. This was described in \citet{Samsing2019}, in which it was suggested that the orbital-period
distribution of the BBH population can be probed using EM observations of stars disrupted by this population.
As described in \citep[e.g.][]{Liu:2009fl, 2014ApJ...786..103L, 2016MNRAS.458.1712R, 2017MNRAS.465.3840C, 2018arXiv181201118L, Samsing2019}, the time dependent EM signal from a TDE involving a BBH will be different from
the standard $t^{-5/3}$ luminosity decay found in the single BH case \citep{Rees1988}, as the second BH now can interact with the tidal stream. This interaction leads to periodic variations in the light curve, including gaps and interruptions, with a period that directly relates to the BBH orbital period. This has been illustrated in the SMBH case using numerical techniques \citep[e.g.][]{Liu:2009fl}, and one candidate has proposed (TDE J1201+30), from which the authors were able to put constraints on the binary orbital period \citep{2014ApJ...786..103L}. Finally, as described in \citep{Samsing2019}, the orbital period distribution of BBH disrupters 
in stellar clusters is very different from other binary distributions, including the well known
Opik's distribution \citep{1924PTarO..25f...1O}, and the distribution that follows from pure GW decay \citep[e.g.][]{2016PhRvL.116w1102S, 2017MNRAS.469..930C}. This indicates that the cluster BBH population can be distinguished from other possible scenarios.

\begin{figure}
\begin{center}
\includegraphics[width=0.5\textwidth]{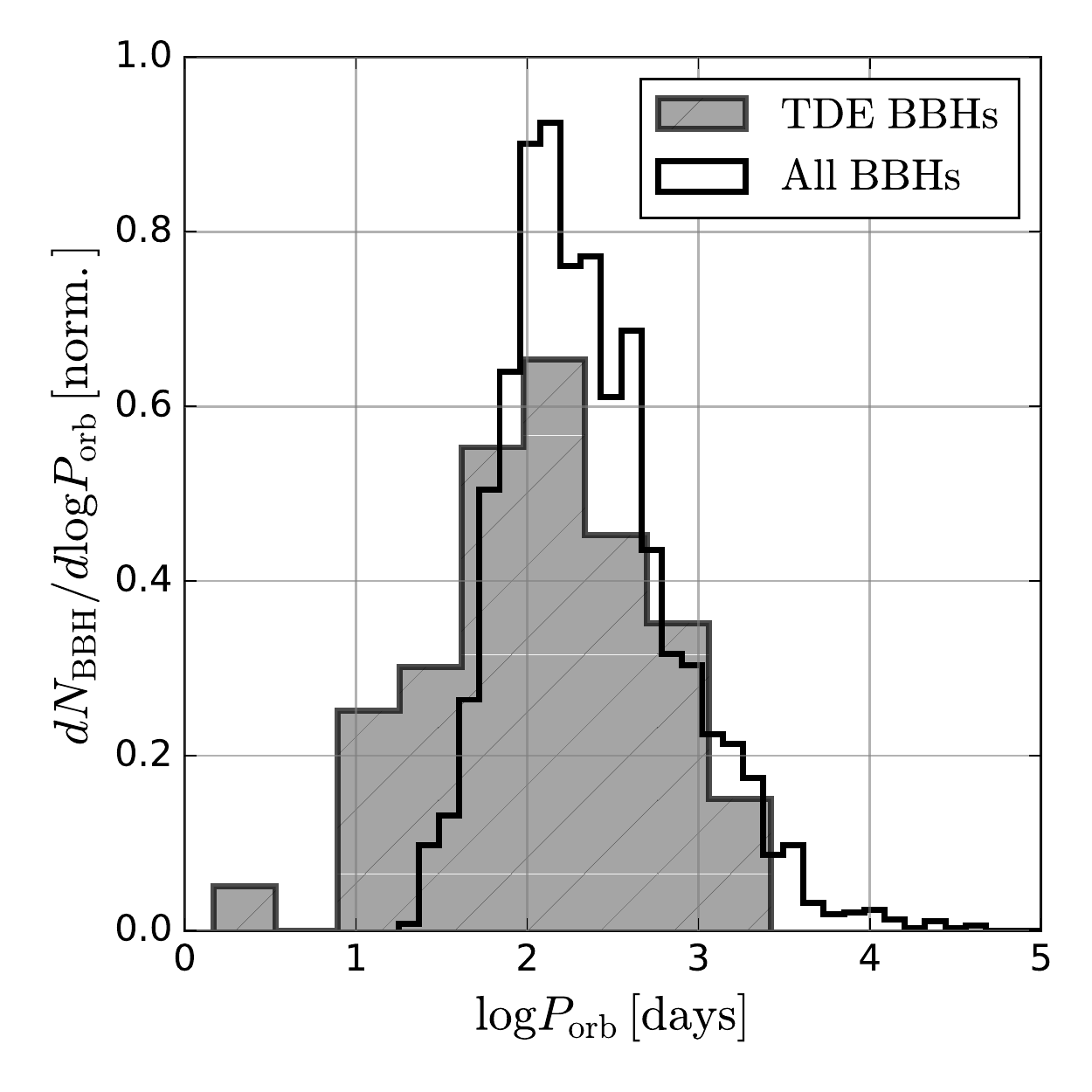}
\caption{\label{fig:Porb} Orbital period distribution for all BBHs that undergo planet TDEs during resonant encounters (hatched gray histogram) compared to orbital period distribution for all BBHs retained in the cluster at representative snapshots in time (black histogram).
}
\end{center}
\end{figure}

In Figure \ref{fig:Porb} is shown the orbital period distribution of all BBHs that lead to planet TDEs during resonant encounters compared to the orbital period distribution for all BBHs retained in the cluster at representative snapshots in time (i.e., snapshots with $t>8\,$ Gyr, representative of old GCs). As the figure shows, the two distributions share several similarities, which is a direct result of that the tidal disruption cross section is independent of the BBH semi-major axis as described in the paragraphs above and as in \citet{Samsing2019}. On the basis of the overlap of these two distributions, figure \ref{fig:Porb} demonstrates that if BBH--planet TDEs are observed in significant numbers, they may be used to indirectly constrain the overall BBH orbital period distribution.

\section{Other Observational Prospects}
\label{sec:other}
\subsection{Interactions of planets with neutron stars and white dwarfs}
\label{sec:NSplanets}

In addition to interacting with BHs, planets also readily mix with both neutron stars (NSs) and white dwarfs (WDs) in typical GCs and through these dynamical interactions, planetary systems with NS/WD host stars will naturally form. The dynamical formation of NS/WD planetary systems in GCs is motivated observationally by the millisecond pulsar (MSP) planet detected through radio observations in the Milky Way cluster M4, which is now thought to be a member of a triple system where the third body is a WD \citep[e.g.,][]{Thorsett1999}.

In Table \ref{table:WDNS}, we list the average number of dynamically-formed NS-- and WD--planet binaries found in each cluster model at any given late-time cluster snapshot ($t$ in range 8--12 Gyr). As shown in the table, WD--planet binaries are more common than NS--planets. This is simply due to the relatively large number of WDs versus NSs (typically roughly $5\times10^4$ versus 500 at late times, as shown in Figure \ref{fig:radial_dist}). Once formed, the NS--planets and WD--planets typically survive for 100s of Myr in the cluster before being broken apart by subsequent dynamical encounters. This is in contrast to the BH--planet systems discussed in Section \ref{sec:BHplanets}, which typically survive for only $\approx10$ Myr before being broken apart. BH planetary systems are expected to have shorter lifetimes than the WD/NS counterparts, simply because BH--planets are preferentially found closer to the cluster's high-density core owing to the high mass of the BH. Furthermore, higher mass means a larger cross section and thereby a shorter interaction timescale ($t_{\rm{int}} \propto M^{-1}$).

As shown in the table, we expect up to roughly 5 NS--planets and up to roughly 100 WD--planets to be found in a GC at any given point in time. As Table \ref{table:WDNS} shows, the exact number depends upon the initial number of planets and on the initial planet orbital properties. In general, the higher the number of planets, the higher the number of WD--planet binaries. For the NS--planets, the trend is less clear, although this is likely simply due to the uncertainties associated with the small number statistics. For planet-to-star ratios higher than the maximum of 0.5 considered here, the numbers of WD--planets and NS--planets may increase even further.

These numbers are also dependent on the various cluster parameters considered. We consider here GCs with present day-mass of roughly $2\times 10^5\,M_{\odot}$, which is typical of the GCs observed in the Milky Way, however in more massive and dense (i.e., core-collapsed) clusters such as M15, a larger number of NS-- and WD--planet systems may be expected due to both the higher number of objects and the higher central density (and therefore higher formation rate) in these clusters. On the other hand, especially for WD--planets, some fraction of which may actually be formed through the stellar evolution of a planet's primordial host star, a relatively dense cluster may in fact lead to a \textit{decrease} in the number of planetary systems as these primordial planetary systems are broken. We leave a more detailed of these possibilities for a future study.

It is well-understood both observationally \citep[e.g.,][]{Ransom2008} and theoretically \citep[e.g.,][]{Ye2018} that MSPs form at a higher rate per unit mass in GCs relative to the Galactic field as a result of dynamical encounters. If some fraction of the NSs that obtain planet companions are in fact MSPs (having been spun up through accretion of material from a companion earlier in the NS's evolutionary history), radio observations of these objects may allow the presence of the planet companion to be inferred. We reserve a more detailed study of the interaction of pulsars with planets in GCs for a future study.

Planets bound to WDs can in principal be detected through standard transit methods with the caveat that WDs are intrinsically faint, so transit searches are limited to smaller distances relative to main-sequence star planetary systems. Recent studies \citep{Zuckerman2010,Koester2014} have shown that roughly a quarter to half of WDs in the Milky Way are observed to have contaminated atmospheres possibly from the accretion of planetary material, indirectly suggesting remnants of planetary systems. Furthermore, recent studies \citep[e.g.,][]{Vanderburg2015,Manser2019} have identified transiting planetesimals around WD hosts. Future surveys such as LSST are likely to provide further insight into the numbers of WD planetary systems in both GCs and the Galactic field \citep[e.g.,][]{Cortes2019}.

\begin{deluxetable}{lc|c|c}
\tabletypesize{\scriptsize}
\tablewidth{0pt}
\tablecaption{Average number of WD-- and NS--planet binaries in all model GCs at late times}
\tablehead{
	\colhead{Model} &
    \colhead{$N_{\rm{P,\,i}}$}&
    \multicolumn{1}{c}{WD--P binaries}&
    \multicolumn{1}{c}{NS--P binaries}\\
}
\startdata
1 & $10^5$        &  18.5 & 1.1 \\
2 & $2\times10^5$ & 42.0 & 0.1\\
3 & $3\times10^5$ &  46.6 & 4.4\\
4 & $4\times10^5$ &  59.9 & 2.4 \\  
\hline
$5^{\alpha}$ & $4\times10^5$ &  73.4 & 0.5\\
$6^{\beta}$ & $2\times10^5$ &  27.0 & 1.6\\
\enddata
\tablecomments{\label{table:WDNS} Models are labeled as in Table \ref{table:models}.}
\end{deluxetable}

\subsection{Ejected free-floating planets}
\label{sec:free-floating}

Theories of planet formation predict the existence of free-floating planets which are expected to have formed in association with a host star\footnote{It has also been proposed that star formation processes may extend down to masses as low as $\sim$a Jupiter mass \citep[e.g.,][]{Boyd2005}, which would allow free-floating planets to also form through fragmentation of gas clouds in a manner similar to standard star formation.} before being unbound through a variety of potential mechanisms including planet-planet interactions \citep[e.g.,][]{Rasio1996,Marzari2002,Chatterjee2008,Veras2009}, post-main sequence evolution of their host star \citep[e.g.,][]{Kratter2012,Veras2016}, or dynamical interactions within a stellar cluster such as those considered here and previously in e.g., \citet{HurleyShara2002,Spurzem2009}.

If a free-floating planet is very closely aligned with a distant source star, the planet can potentially be identified through gravitational microlensing by observing a transient brightening of the source star caused by the focusing of light rays of the source by the gravitational field of the planet. A number of ground-based surveys such as the Microlensing Observations in Astrophysics \citep[MOA-II;][]{Sumi2003}, the Optical Gravitational Lensing Experiment \citep[OGLE-IV;][]{Udalski2015} and the Korean Microlensing Telescope Network \citep[KMTNet;][]{Kim2016} should in principal be capable of detecting such events. Furthermore, in the coming years, WFIRST is expected to identify in excess of roughly 1000 free-floating planets in the Milky Way through microlensing \citep{WFIRST2015}. 

As shown by columns 5-7 of Table \ref{table:models} as well as Figure \ref{fig:cluster_properties}, a large number of planets are ejected by their host cluster into the Galactic halo as a result of dynamical encounters and tidal loss.
These planets can be ejected both as binaries (i.e., still bound to a host star) or as single planets. Scaling our models to the Milky Way GC population as in Section \ref{sec:EM}, we predict up to as many as $10^7$ planet--star systems and $2\times10^7$ single planets may presently populate the Galactic halo following dynamical ejection from GCs. 

Although the number of free-floating planets produced through cluster dynamics is likely small compared to the number of free-floating planets produced through standard planet-planet dynamics for planetary systems in the field \citep[recent studies predict as many as 1-2 free-floating planets may exist for every star in the Galactic field;][]{WFIRST2015}, several features of these cluster-specific planets may potentially allow them to be distinguished from their Galactic-field counterparts. For instance, 
free-floating planets that originate in the Galactic field will preferentially be located in the plane of the Galaxy, in contrast to the planets ejected from clusters which will adopt the spherically-isotropic distribution of their host clusters. Thus, cluster-born planets may constitute a relatively large fraction of free-floating planets which are identified in the Galactic halo. We reserve a focused analysis exploring potentially detectability of free-floating planets for a later study.

\section{Conclusions and Discussion}
\label{sec:Discussion}

\subsection{Summary}

Using H\'{e}non-type Monte Carlo cluster models, we have studied the evolution of planetary systems in dense star clusters. We briefly summarize our main findings below:
\begin{enumerate}

\item We show that a large fraction of primordial planetary systems in a cluster are broken apart through dynamical encounters with other objects. Assuming an initial orbital separation of 1 AU, we find roughly $10\%$ of all primordial planetary systems are broken and assuming an initial orbital separation of 5 AU, we find this percentage can reach as high as roughly $50\%$.
\item Furthermore, a large number ($30\%-80\%$, depending on initial conditions) of primordial planets are ejected from their host cluster through dynamical encounters and tidal loss, either as single planets or still bound to a host star.
\item Both single planets and star--planet binaries naturally mix with stellar-mass BHs in the central regions of their host cluster, leading to the dynamical formation of BH--planet binaries throughout the cluster lifetime.
\item As a secondary consequence of the dynamical mixing with the cluster's BH population, planets will frequently pass close enough to BHs to be tidally disrupted. These BH--planet TDEs will likely produce flares with characteristic emission timescales of roughly days and peak luminosities of a few $\times 10^{40}\,\rm{erg\,s}^{-1}$.
\item  These BH--planet TDEs may be detectable by LSST at a rate of roughly a few events per year or higher. If observed, these TDEs may place further constraints on both planet and BH populations in GCs.
\item Finally, we have shown that up to a few NS--planet binaries are expected to be found in typical clusters. If their NS hosts have been spun up to become radio millisecond pulsars through previous interactions, these planet companions could be identified through radio observations. 
\end{enumerate}

In principle, TDEs of planets may also occur through interaction with WDs and NSs. \citet{DelSanto2014} noted that a peculiar transient event observed in the Milky Way GC NGC 6388 has features consistent with a planet tidally disrupted by a massive WD. We identify roughly 500 WD--planet TDEs in our set of cluster models (or roughly $10^{-8}\,\rm{yr}^{-1}$ per cluster), a factor of roughly a few times lower than BH--planet TDEs. Although WDs can outnumber BHs significantly, especially at late times (see Figure \ref{fig:radial_dist}), the TDE rate for WDs is limited by the relatively low mass of a WD compared to a BH (roughly $0.7 M_{\odot}$ compared to $20 M_{\odot}$): as shown in Equation \ref{eq:TDE}, $\Gamma_{\rm{TDE}} \propto M^{4/3}$, where $M$ is the mass of the disrupting object. A detailed examination of WD--planet TDEs is beyond the scope of our study. However, we note that the potential detection described in \citet{DelSanto2014} points toward the possibility of planet TDEs in GCs and, in particular, given the fact that BH--planet TDEs are expected to outnumber WD--planet TDEs by a factor of a few, further suggests that upcoming transient surveys like LSST may indeed observe such events.

The rate of WD--planet TDEs suggested by our models is signficantly lower than the rate of $3.3\times10^{-6}\,\rm{yr}^{-1}$ quoted by \citet{DelSanto2014}, but is roughly consistent given that \citet{DelSanto2014} assumed a relatively high planet-to-star ratio ($F_{\rm{p}} = 10-100$) compared to our models ($F_{\rm{p}}=0.1-0.5$). Indeed, as pointed out by \citet{DelSanto2014} the detection of planet TDEs linked to clusters may place contraints on the total number of planets (both single and bound to host stars) within GCs.

\subsection{Directions for future work}
\label{sec:future}

As this paper is a first attempt at exploring the mixtures of planets with stellar remnants in the context of the \texttt{CMC} cluster dynamics code, a number of simplifying assumptions have been made. A more detailed study of several of these assumptions may form the basis for future work on the topic. For one, a future study may implement a more realistic planet mass spectrum which extends beyond Jupiter-like planets to masses down to Earth-mass or lower. Although the inclusion of lower mass planets is unlikely to strongly affect the dynamical influence of planets on their host cluster, lower mass planets may lead to different types of TDEs. In particular, as discussed in, e.g., \citet{Gourg2019}, Earth-like planets have lower densities than their Jupiter-like counterparts, and therefore must pass relatively close to a stellar remnant to reach Roche or tidal contact. As a result, the probability of undergoing a TDE is likely lower for an Earth-like planet than a Jupiter-like planet. Furthermore, because of the differences in their chemical composition, TDEs of lower-mass planets may yield different types of electromagnetic signatures.

This analysis considered the disruption of planetary systems by dynamical encounters with other objects in their host cluster. However, it is well understood that internal planetary dynamics and other mechanisms (see Section \ref{sec:free-floating}) will also unbind planets from their host star. Therefore, prior to any cluster-specific dynamical interactions a large population of single planets may already have been created. In this case, it may be appropriate to also include a population of single planets within the cluster initial mass function. Indeed, the single planets may even outnumber the stars by a factor of up to $\sim100$ \citet{HurleyShara2002}. These single planets will also be susceptible to interactions with stellar remnants and thus the inclusion of these objects may lead to an increase in the remnant--planet TDE rate by a potentially significant amount.

Here we have focused exclusively on lower-metallicity clusters ($Z=0.1\,Z_{\odot}$) with the motivation of exploring clusters that resemble the low-metallicity old GCs observed in the Milky Way. However, planet formation (as well as a number of other cluster features, including the mass spectrum of BH remnants) depends in a significant way upon metallicity \citep[e.g.,][]{Fischer2005,Brewer2018}. Therefore an exploration of metallicities and ages representative of the young massive clusters observed in the local universe \citep[see, e.g.,][]{PortegiesZwart2010} presents another relevant and interesting avenue for study.

Finally, we have limited the present study to cluster models of a fixed initial number of stars ($N=8\times 10^5$). Although this particular choice is representative of an ``average'' GC in the Milky Way, we fail to capture the specific details that may be relevant for lower or higher mass clusters. Furthermore, we have considered for simplicity only clusters with initial virial radii of 1 and 2 pc, however, previous analyses \citep[e.g.,][]{Kremer2019a} have demonstrated that a wider range in initial cluster size is necessary to produce the full spectrum of cluster types observed in the Milky Way at present. For more massive and dense clusters (extending up to nuclear star clusters), the rates of various processes discussed in this work (dynamical disruption of planetary systems, formation of planet--compact object binaries, and BH--planet TDEs) may be amplified considerably. We hope to explore the evolution of planetary systems in a wider array of cluster types, which would include clusters of different initial mass and size, in a future study.

\acknowledgments
We thank Steinn Sigurdsson and Enrico-Ramirez Ruiz for useful discussions. This work was supported by NASA ATP Grant NNX14AP92G and NSF Grant AST-1716762. KK acknowledges support by the National Science Foundation Graduate Research Fellowship Program under Grant No. DGE-1324585. JS acknowledges support from the Lyman Spitzer Fellowship.
DJD acknowledges financial
support from NASA through Einstein Postdoctoral Fellowship award
number PF6-170151.

\bibliographystyle{aasjournal}
\bibliography{mybib}

\listofchanges

\end{document}